\documentstyle[12pt]{report}
\def\PRL #1 #2 #3{{\sl Phys. Rev. Lett.} {\bf#1} (#2) #3}
\def\NPB #1 #2 #3{{\sl Nucl. Phys.} {\bf B#1} (#2) #3}
\def\NPBFS #1 #2 #3 #4{{\sl Nucl. Phys.} {\bf B#2} [FS#1] (#3) #4}
\def\CMP #1 #2 #3{{\sl Commun. Math. Phys.} {\bf #1} (#2) #3}
\def\PRD #1 #2 #3{{\sl Phys. Rev.} {\bf D#1} (#2) #3}
\def\PLA #1 #2 #3{{\sl Phys. Lett.} {\bf #1A} (#2) #3}
\def\PLB #1 #2 #3{{\sl Phys. Lett.} {\bf #1B} (#2) #3}
\def\JMP #1 #2 #3{{\sl J. Math. Phys.} {\bf #1} (#2) #3}
\def\PTP #1 #2 #3{{\sl Prog. Theor. Phys.} {\bf #1} (#2) #3}
\def\SPTP #1 #2 #3{{\sl Suppl. Prog. Theor. Phys.} {\bf #1} (#2) #3}
\def\AoP #1 #2 #3{{\sl Ann. of Phys.} {\bf #1} (#2) #3}
\def\PNAS #1 #2 #3{{\sl Proc. Natl. Acad. Sci. USA} {\bf #1} (#2) #3}
\def\RMP #1 #2 #3{{\sl Rev. Mod. Phys.} {\bf #1} (#2) #3}
\def\PR #1 #2 #3{{\sl Phys. Reports} {\bf #1} (#2) #3}
\def\AoM #1 #2 #3{{\sl Ann. of Math.} {\bf #1} (#2) #3}
\def\UMN #1 #2 #3{{\sl Usp. Mat. Nauk} {\bf #1} (#2) #3}
\def\FAP #1 #2 #3{{\sl Funkt. Anal. Prilozheniya} {\bf #1} (#2) #3}
\def\FAaIA #1 #2 #3{{\sl Functional Analysis and Its Application} {\bf
#1} (#2) #3}
\def\BAMS #1 #2 #3{{\sl Bull. Am. Math. Soc.} {\bf #1} (#2)
#3} \def\TAMS #1 #2 #3{{\sl Trans. Am. Math. Soc.} {\bf #1} (#2) #3}
\def\InvM #1 #2 #3{{\sl Invent. Math.} {\bf #1} (#2) #3}
\def\LMP #1 #2 #3{{\sl Letters in Math. Phys.} {\bf #1} (#2) #3}
\def\IJMPA #1 #2 #3{{\sl Int. J. Mod. Phys.} {\bf A#1} (#2) #3}
\def\AdM #1 #2 #3{{\sl Advances in Math.} {\bf #1} (#2) #3}
\def\RMaP #1 #2 #3{{\sl Reports on Math. Phys.} {\bf #1} (#2) #3}
\def\IJM #1 #2 #3{{\sl Ill. J. Math.} {\bf #1} (#2) #3}
\def\APP #1 #2 #3{{\sl Acta Phys. Polon.} {\bf #1} (#2) #3}
\def\TMP #1 #2 #3{{\sl Theor. Mat. Phys.} {\bf #1} (#2) #3}
\def\JPA #1 #2 #3{{\sl J. Physics} {\bf A#1} (#2) #3}
\def\JSM #1 #2 #3{{\sl J. Soviet Math.} {\bf #1} (#2) #3}
\def\MPLA #1 #2 #3{{\sl Mod. Phys. Lett.} {\bf A#1} (#2) #3}
\def\JETP #1 #2 #3{{\sl Sov. Phys. JETP} {\bf #1} (#2) #3}
\def\JETPL #1 #2 #3{{\sl  Sov. Phys. JETP Lett.} {\bf #1} (#2) #3}
\def\PHSA #1 #2 #3{{\sl Physica} {\bf A#1} (#2) #3}
\def\CQG #1 #2 #3{{\sl Class. Quantum Grav.} {\bf #1} (#2) #3}
\def\SJNP #1 #2 #3{{\sl Sov. J. Nucl. Phys. (Yadern.Fiz.)} {\bf #1} (#2) #3}
\def\a{\alpha}\def\b{\beta}\def\g{\gamma}\def\d{\delta}\def\e{\epsilon}

\def\k{\kappa}\def\s{\sigma}
\def\Th{\Theta}\def\om{\omega}\def\Om{\Omega}\def\G{\Gamma}

\newcommand{\nn}{\nonumber\\}\newcommand{\p}[1]{(\ref{#1})}
\begin{document}
\renewcommand{\thefootnote}{\fnsymbol{footnote}}
\thispagestyle{empty}
\begin{flushright}
Prerint DFPD 95/TH/92
\end{flushright}

\medskip
\begin{center}
{\large\bf
SUPERSTRINGS AND SUPERMEMBRANES
IN THE DOUBLY SUPERSYMMETRIC
GEOMETRICAL APPROACH}
\vspace{0.3cm}

{\bf Igor A. Bandos,}
\footnote{Supported  in  part  by  the
International Science Foundation under the grant N RY 9000,
 by the State  Committee  for  Science  and  Technology  of
Ukraine under the Grant N 2/100 and
 by the INTAS grants 93--127, 93--493, 93--633}

\vspace{0.2cm}
{\it Kharkov Institute of Physics and Technology}
{\it 310108, Kharkov,  Ukraine}\\
e-mail:  kfti@kfti.kharkov.ua

\vspace{0.2cm}
\renewcommand{\thefootnote}{\dagger}
{\bf Dmitrij Sorokin\footnote{on leave from Kharkov Institute of
Physics and Technology, Kharkov, 310108, Ukraine.\\e--mail:
sorokin@pd.infn.it}$^*$,
\renewcommand{\thefootnote}{\ddagger}
Mario Tonin
\footnote{e--mail: tonin@pd.infn.it},
\renewcommand{\thefootnote}{\fnsymbol{footnote}\fnsymbol{footnote}}
\footnote{Supported in part by M.P.I. Work carried out as part of the
European Community Programme ``Gauge Theories, Applied Supersymmetry and
Quantum Gravity'' under contract SC1--CT92--D789.}
Paolo Pasti$^{\fnsymbol{footnote}\fnsymbol{footnote}}$}

\vspace{0.2cm}
{\it Universit\`a Degli Studi Di Padova
Dipartimento Di Fisica ``Galileo Galilei''\\
ed INFN, Sezione Di Padova
Via F. Marzolo, 8, 35131 Padova, Italia}

\vspace{0.2cm}
{\bf and}

\vspace{0.2cm}

{\bf Dmitrij V. Volkov}$^*$

\vspace{0.2cm}
{\it Kharkov Institute of Physics and Technology}
{\it 310108, Kharkov,  Ukraine}\\
e-mail:  dvolkov@kfti.kharkov.ua

\vspace{0.5cm}
%\vspace{0.3cm}
{\bf Abstract}
\end{center}

We perform a generalization of the geometrical approach to
describing extended objects for studying the doubly supersymmetric
twistor--like formulation of super--p--branes. Some basic features of
embedding world supersurface into target superspace specified by a
geometrodynamical condition are considered. It is shown
that the main attributes of the geometrical approach, such as the second
fundamental form and extrinsic torsion of the embedded surface, and
the Codazzi, Gauss and Ricci equations,  have their doubly
supersymmetric counterparts. At the same time the embedding of
supersurface into target superspace has its particular features. For
instance, the embedding may cause more rigid
restrictions on the geometrical properties of the supersurface. This is
demonstrated with the examples of an N=1 twistor--like
supermembrane in D=11 and type II superstrings in D=10,
where the geometrodynamical condition causes the embedded
supersurface to be minimal and puts the theories on the mass shell.

\setcounter{page}1
\renewcommand{\thefootnote}{\arabic{footnote}}
\setcounter{footnote}0

\newpage
{\large \bf Introduction}

\bigskip
Finding the most adequate way to describe physical objects is an important
problem which, very often, allows one to achieve deeper knowledge and
perform further development of the corresponding theory.
One of the most typical examples is the theory of strings and
superstrings, various formulations of which throw light on different
features of the string.

Among the string formulations there is a so called geometrical approach,
which is essentially based on the theory of surfaces embedded into a
target space. This approach was originated in papers by Lund and Regge
\cite{lr} and Omnes \cite{om}, and revealed a connection of the string
equations of motion with two--dimensional (exactly solvable) non--linear
equations, such as the sin--Gordon and Liouville equation.

Though, of course, all string formulations imply that string world--sheet
is a surface embedded into a target space--time, the geometrical approach
explores this in the most direct way by dealing with such objects as a
target--space moving frame at every point of the surface, extrinsic
curvature and torsion of the surface, and reducing the string equations to
the system of the Codazzi, Gauss and Ricci equations completely
determining the embedding of the surface.

The geometrical approach was studied in connection with the problem of
formulating consistent quantum string theory in non--critical space--time
dimensions and has been developed in application to strings and
$p$--branes in a number of papers (see \cite{barnes,zhelt} and
references therein).

The interest to the approach is due to the deep
relationship of $p$-brane equations of motion with equations describing
non--linear systems such as $\sigma$--models in
${ SO(1,D-1) \over {SO(1,p) \times SO(D-p-1)}}$ target space, and exactly
solvable and completely integrable dynamical systems (in the case of
strings) \cite{lr,om,barnes,zhelt}. In particular, it is remarkable that
choosing a Lorentz--covariant gauge, one can reduce the number of the
string coordinates in a $D$--dimensional space--time to $(D-2)$
independent variables subject to a system of non--linear
differential equations \cite{barnes,zhelt} for which the general solution
can be constructed.

As to membranes, a relation between their non--linear equations of motion
and that of integrable systems has been found as well \cite{hoppe}.
And since the problem of complete solving the membrane equations of motion
is still open, the attempts to reformulate membrane theory directly in the
geometrical framework of surface theory seem to be justified. Using
this approach one may hope to find new physically interesting solutions
%(at least particular ones)
to the membrane equations, and gain deeper
insight into the problem of string--membrane duality \cite{m1c,m1d,m2}.

To develop the geometrical approach,
in addition to $p$--brane space--time coordinates
one introduces
auxiliary world surface fields
describing a target space moving frame attached to every point of the
world surface,
so that a system of equations specifying the parallel transport of the
moving frame along the world surface is equivalent to the
$p$--brane equations of motion. This determines a geometry on the
surface induced by embedding. Note that the intrinsic geometry of the
world surface (and the corresponding part of the induced geometry)
characterize internal properties of the surface and the local gauge
symmetries of the model, while the extrinsic part of the induced
geometry specifies the motion of the p--brane in the target space.

Moving frame components can be introduced
directly into a $p$--brane action
\footnote{
$P$--brane models of this kind have been
considered, for example, in \cite{vzUFZh,zhelt1}.
}
the latter being
considered as a dynamical ground for the geometrical approach
\cite{lr,om,barnes}.

 Here the question arises what is the natural way for introducing the
moving frame into the super--$p$--brane theory \cite{gschw84}--\cite{m1a}.

One of the possibilities is a twistor--like formulation of
super--$p$--branes \cite{fer}--\cite{bers94}
which provided the geometrical solution
\cite{stv} to the problem of local fermionic $\kappa$--symmetry
\cite{al,sig,gsw}.

The twistor--like formulation is based on a notion of double
supersymmetry originally introduced for constructing more general
supersymmetric models \cite{spinsup} studied, in particular, in
connection with the problem of coupling worldsheet supergravity to target
space supergravity for unique treatment of the Neveu--Schwarz--Ramond and
Green--Schwarz superstrings \cite{vz,stvz,to}.

In the doubly supersymmetric formulation of super--$p$--branes auxiliary
commuting spinor variables, having properties of twistors
\cite{pen}--\cite{stv} and Lorentz harmonics
\cite{gikos},\cite{sok}--\cite{kr,bh,bz0,ghs,gds,bzst,bzm} appear
as superpartners
of the target
superspace Grassmann coordinates, their bilinear combinations  forming
Lorentz vectors which can be identified  with components of local moving
frame in the target superspace. This provides the ground for a
generalization and a development of the geometrical approach,
which implies studying the embedding of a world
supersurface into a target superspace.

In the present paper we perform the first steps in this direction and
consider as examples an $N=1$ supermembrane in $D=11$ and
superstrings in $D=10$.

In the doubly supersymmetric formulation of super--$p$--branes the
embedding of a world supersurface into a target superspace is specified by
a geometrodynamical condition (see section 2.2), which prescribes the
pullback of a target superspace one--form onto the world supersurface to
have zero components along Grassmann directions of the latter
\cite{stv}--\cite{bers94}.
The twistor--like solution to the Virasoro constraints arises as an
integrability condition for the geometrodynamical equation. In the case of
the D=11, N=1 supermembrane and the D=10, N=2 superstring imposing the
geometrodynamical condition puts the theory on the mass shell, which
causes the problem with constructing worldsheet superfield actions, as was
noticed by Galperin and Sokatchev \cite{gs2}.

Below, when considering the
doubly supersymmetric p--branes we will not discuss the problem of getting
the action,
since for our purpose of developing the geometrical approach just the
equations of motion of super--$p$--branes are required. So for the two
theories under consideration the geometrodynamical condition can be
regarded as one determining a minimal supersurface in a target superspace,
and we will use it  as the starting point for getting geometrical
equations analogous to the Codazzi, Gauss and Ricci equation.

Leaving apart the problem of constructing the superfield action
we also will not touch one important ingredient of the
super--p--branes in the Green--Schwarz \cite{gschw84,gsw} as well as the
twistor--like superfield \cite{hsstr}--\cite{bers94}
 formulation, namely a Wess--Zumino term
and a corresponding Wess--Zumino differential form. In the Green--Schwarz
formulation the crucial role of the Wess--Zumino term in the action is to
ensure the local fermionic $\k$--symmetry. In the twistor--like action,
in addition to the geometrodynamical term, the pullback of the
Wess-Zumino form further specifies the embedding of the world supersurface
and generates super--p--brane tension, thus turning a null super--p--brane
\cite{bstv} into the valuable extended object \cite{dghs92}. The
Wess--Zumino term
is a differential form on the world supersurface which is a closed form on
the mass shell provided the geometrodynamical condition takes place
(see Tonin in \cite{hsstr}, and \cite{dghs92}), and when one gets the
equations of motion from a
super--p--brane
action they contain the contribution from the Wess--Zumino term. Thus,
as soon as the equations of motion are obtained (for instance, as a
consequence of the geometrodynamical condition)
 the Wess--Zumino term does not provide any new
information.

The paper is organized as follows.

In Chapter 1 we review the main features of the geometrical approach to
bosonic p--branes by introducing the notion of the local frame,
presenting an appropriate p--brane action to start with and rewriting
the p--brane equations of motion in the form of the Codazzi, Gauss and Ricci
equations for the second fundamental form and extrinsic torsion of world
surface embedded into target space--time.

In Chapter 2 we perform a generalization of the geometrical approach to
the case of super--p--branes. It is shown that the basic role in the
formulation is played by a spinor local frame in target superspace
the local vector frame being composed of the spinor one. The
embedding of world supersurface is specified by the geometrodynamical
condition. The supersymmetric analogues of the Codazzi, Gauss and Ricci
equations and of the second fundamental form are considered. A condition
for the embedded supersurface to be minimal is found.

In Chapter 3 and 4 the results of Chapter 2 are applied for studying
particular features of D=11 N=1 supermembranes and D=10 type II
superstrings, and it is shown that world supersurface embedding
specified by the geometrodynamical condition is minimal in contrast to
the case of a heterotic string.

In Conclusion we sum up the results obtained.

Our notation and convention are as follows. The small Latin indices stand
for vectors and the Greek indices stand for spinors. All underlined
indices correspond to target (super)space of D bosonic dimensions, and
that which are not underlined correspond to world (super)surface of
(p+1) bosonic dimensions. The indices from the beginning of the
alphabets denote the vector and spinor components in the tangent
(super)space. Indices from the second half of the alphabets are world
indices:
$$
{\underline a},{\underline b},{\underline c}=0,...,D-1\qquad
{\underline l},{\underline m},{\underline n}=0,...,D-1;
$$
$$
a,b,c=0,...,p \qquad l,m,n=0,...,p
$$
$$
{\underline \a},{\underline \b},{\underline \g}=1,...,2^{[{D\over 2}]}
{}~~(or~2^{[{D\over 2}-1]})\qquad
{\underline \mu},{\underline \nu},{\underline \rho}=1,...,2^{[{D\over 2}]}
{}~~(or~2^{[{D\over 2}-1]});
$$
$$
\a,\b,\g=1,...,2^{[{p\over 2}]}\qquad
\mu,\nu,\rho=1,...,2^{[{p\over 2}]}
$$
$
i,j,k=1,...,D-p-1$~~{\rm{stand~for~ the ~vector ~representation~ of}~
SO(D-p-1)};\\
$
p,q,r~(or~\dot p,\dot q,\dot r)=1,...,D-p-1$~~~{\rm{stand~for~ a spinor
 ~representation~ of}~ SO(D-p-1)}.

More information about notation and convention the reader may find in
the main text or in the Appendices.

%\section{
\chapter{
Geometrical approach to  bosonic $p$--branes }
\section{Moving frame on the embedded surface}
To describe an embedding of a $(p+1)$--dimensional world surface into a
$D$--dimensional flat space--time one introduces in the target space
(parametrized by $x^{\underline m}$) a local moving frame
$u^{\underline{a}} (x^{\underline{m}}) \equiv dx^{\underline{n}}
u^{~\underline{a}}_{\underline{n}} (x^{\underline{m}}) $,
whose components
$u^{~\underline{a}}_{\underline{n}} (x^{\underline{m}}) $
satisfy the orthonormality conditions

\begin{equation}\label{1.1}
u^{~\underline{a}}_{\underline{m}}\eta^{\underline{m}\underline{n}}
u_{\underline{n}\underline{b}} = \eta ^{\underline{a}\underline{b}}=
{\it diag}(1,-1,\ldots ,-1)
\end{equation}
Eq. \p{1.1} restricts the matrix
$|| u^{~\underline{a}}_{\underline{n}} (x^{\underline{m}}) ||$
to take its values in the Lorentz group \mbox{$SO(1,D-1)$}. Thus, in
particular,
using an appropriate Lorentz  transformations in the tangent space,
one can always choose a frame whose components form the unit matrix
$\d^{\underline{a}}_{\underline{m}}$ globally in the flat space.

On the other hand, by use of the Lorentz transformations one can adjust a
local frame to the embedded surface
$x^{\underline{m}} = x^{\underline{m}} (\xi^{{m}})~$
in such a way that $d=(p+1)$ vectors $u^{~a}_{\underline{m}}$ of the frame
are parallel and $(D-p-1)$ vectors
$u^{~i}_{\underline{m}}$ are orthogonal to the surface at each point of
the latter:
\begin{equation}\label{1.3}
\partial_{m} x^{\underline{m}} u^{~i}_{\underline{m}} = 0 \qquad
\end{equation}
\begin{equation}\label{1.4}
\partial_{m} x^{\underline{m}} u^{~a}_{\underline{m}}
{}~~~\sim ~~~
e^{~a}_{m} ,
\end{equation}
where
$e^{~a} \equiv d \xi^{m} e^{~a}_{m} (\xi)$
is an intrinsic vielbein form on the surface (locally parametrized by
$\xi^m$).  Eqs. \p{1.3}, \p{1.4} determine the moving frame
$u^{~\underline{a}}_{\underline{m}} $ up to a local transformations of
the subgroup $SO(1,p) \times SO(D-p-1)$ of the Lorentz group, SO(1,p)
being identified with the structure group of the surface.  Thus,
$u^{~\underline{a}}_{\underline{m}} $ can be regarded as Lorentz
harmonics parametrizing the coset space ${SO(1,D-1) \over {SO(1,p)
\times SO(D-p-1)}}$.

If we consider a $d=p+1$ -- dimensional surface as one created by a
$p$--brane moving in space--time, Eqs. \p{1.3}, \p{1.4} should be regarded
as $p$--brane equations of motion derived from the action principle.
An appropriate action is \cite{bzst,bzst1,bzm,bzm1}
\begin{equation}\label{2.1}
S_{{ D,p}} = (\a ')^{-{1 \over 2}}
\int d^{{ p+1}} \xi \hskip9pt e(\xi )
\left( -  e^{m}_{a} \partial_{m} x^{\underline m}
u^{a}_{{\underline m}} + p \right) ,
\end{equation}
or in the language of differential forms:
\begin{equation}\label{2.1a}
S_{ D,p} =-{{(-1)^p(\a ')^{-{1 \over 2}}}\over{p!}}
\int u^{a_1}e^{a_2}...
e^{a_{(p+1)}}\varepsilon_{a_1a_2...a_{(p+1)}}+
{{(-1)^p(\a ')^{-{1 \over 2}}p}\over{(p+1)!}}\int
e^{a_1} e^{a_2}... e^{a_{(p+1)}}\varepsilon_{a_1a_2...a_{(p+1)}}.
\end{equation}
Here
 $\alpha^\prime$ is a dimensional constant
 (the Regge slope parameter for the string case $p=1$). The wedge product
 and the exterior derivative of the differential forms are implied in
eq.\p{2.1a} and below where applicable.

Eq.\p{2.1} (or \p{2.1a}) is classically equivalent to the conventional
$p$--brane action \cite{dirac,nambu,polya,howet} (see below).
One can see that the first term in \p{2.1a} differs from the second
(cosmological) term only by the replacement of one of the vielbein forms
$e^a$ with the one--form $u^a=dx^{\underline m}u^a_{\underline
m}$, where $u^{a}_{\underline m}$ are the $d=p+1$ orthonormal vectors from
the moving frame \p{1.1} which transform under $SO(1,p)$. Note that
{\sl a priori} $u^a$
is independent of $e^a$ , and eqs. \p{1.3}, \p{1.4}
arise as the equations of motion of $u_{\underline m}^a$. Since
$u^{a}_{\underline m}$ are world--surface fields subject to \p{1.1}, one
must take into account \p{1.1} when varying \p{2.1} with respect to
$u^{a}_{\underline m}$. This can be performed either by explicit including
the constraint \p{1.1} into the action, or by taking only such variations
of $u^{a}_{\underline m}$, which do not break \p{1.1}. We shall use the
latter procedure which turns out to be more convenient, especially when
dealing with spinor Lorentz harmonics \cite{wieg},
\cite{bh}--\cite{bz0}.

Apparently, the variations of $u^{\underline a}_{\underline m}$
which do not violate \p{1.1} are determined by quantities
$O^{\underline{ab}}=-O^{\underline{ba}}$
taking their values in the algebra of $SO(1,D-1)$:
\begin{equation}\label{2.7}
\d u^{\underline{a}}_{{\underline m}} =
u_{{\underline m} \underline{b}}
O^{\underline{b}\underline{a}}.
\end{equation}

For the differentials of $u^{\underline{a}}_{{\underline m}}$ we get the
same expressions as \p{2.7}:
\begin{equation}\label{2.5}
du^{~{\underline a}}_{{\underline m}} =
u_{{\underline m} \underline b} \Omega^{\underline b\underline a}(d),
\end{equation}
where
$\Omega^{\underline{a}\underline{b}}(d)
=  u^{\underline{a}}_{\underline{m}}
d u^{\underline{b}}_{\underline{m}}
$
are $SO(1,D-1)$ Cartan forms.

For the vectors parallel and orthogonal to the world surface eq.\p{2.5}
splits as follows:
\begin{equation}\label{2.5a}
du^{~{a}}_{{\underline m}} =
u_{{\underline m}b} \Omega^{ba}(d)+u_{{\underline m}k} \Omega^{ka}(d)
\end{equation}
\begin{equation}\label{2.5b}
du^{~{i}}_{{\underline m}} =
u_{{\underline m}b} \Omega^{bi}(d)+u_{{\underline m}k} \Omega^{ki}(d),
\end{equation}
where the one--forms $\Omega^{ab}(d)$ and $\Omega^{ik}(d)$ take their
values in the $SO(1,p)$ and $SO(D-p-1)$ subalgebra of the $SO(1,D-1)$
algebra, respectively, and $\Omega^{ai}(d)$ belong to the cotangent space
of the coset space ${SO(1,D-1) \over {SO(1,p) \times SO(D-p-1)}}$.

By definition $\Omega^{\underline a\underline b}(d)$ satisfy the
Maurer--Cartan equations:
\begin{equation}\label{2.42}
d\Omega ^{ai} -
\Omega ^{a}_{~b} \Omega ^{bi} +
\Omega ^{aj} \Omega ^{ji} = 0,
\end{equation}
\begin{equation}\label{2.43}
{R}^{ab}(d,d)=
d\Omega ^{ab} -
\Omega ^{a}_{~c} \Omega ^{cb} =
\Omega ^{ai} \Omega^{bi},
\end{equation}
\begin{equation}\label{2.44}
{R}^{ij}(d,d)=
d\Omega ^{ij} +
\Omega ^{ij^\prime} \Omega^{j^\prime j} =
\Omega ^{ai} \Omega^{~j}_{a},
\end{equation}
where $d$ is the right external differential
($d(\Om_p\Om_q)=
\Om_pd\Om_q+ (-1)^q d\Om_p\Om_q$ for a product of any p-- and q-- form).

Now we are ready to show the classical equivalence of the
$p$--brane formulation considered and the conventional one
\cite{dirac}--\cite{howet} (see \cite{bzst1} for the string case).

The equations $\d S / \d e^{m}_{a} = 0$ and
$\int \d u^{~a}_{\underline m}  \d S /\d u^{~a}_{\underline m} = 0$ (with $\d
u^{~a}_{\underline m}$
from \p{2.7}) have the form
\begin{equation}\label{2.8}
\partial_{m} x^{\underline m}
u^{~a}_{\underline m} =
e^{~a}_{m},
\end{equation}
\begin{equation}\label{2.9}
\partial_{m} x^{\underline m} u^{i}_{\underline m} = 0,
\end{equation}
in which one can recognize the embedding conditions \p{1.3}, \p{1.4}.
By use of the orthonormality conditions \p{1.1}
eqs. \p{2.8}, \p{2.9} can be rewritten as follows
\begin{equation}\label{2.11}
dx^{\underline m} =
e_{a} u^{{\underline m}a},
\end{equation}
or
\begin{equation}\label{2.12}
u^{{\underline m}a}  =
e^{am} \partial_{m} x^{\underline m},
\end{equation}
Finally, varying \p{2.1} with respect to $x^{\underline{m}}$ one gets
\begin{equation}\label{2.13}
\partial_{m} \left( e e^{~m}_{a} u^{{\underline m}a}
\right) = 0.
\end{equation}

Substituting eq. \p{2.12} back into \p{2.1} and introducing
a world surface metric
$$ g_{mn} \equiv e_{m a} e^{~a}_{n}=
\partial_{m}x^{\underline m} \partial_{n}x_{\underline m}
$$
we derive a conventional action functional for bosonic
$p$--branes \cite{dirac} -- \cite{howet}:
\begin{equation}\label{nambu}
S_{conventional}=
\int d\xi^{(p+1)}
det^{{1/2}}(\partial_{m}x^{\underline m} \partial_{n}x_{\underline
m}).
\end{equation}
In conclusion to this section we would like to draw attention to the
fact that
the bosonic p-brane action in the form \p{2.1} does not possess Weyl
invariance ($e^a_m~\rightarrow~We^a_m$) even for the case of strings (p=1)
(in contrast to the  string action
$\int d\xi^2 eg^{mn}\partial_{m}x^{\underline m} \partial_{n}x_{\underline
m}$ which also involves the intrinsic worldsheet metric). In this
respect \p{2.1} is closer to the Nambu action \p{nambu}. To get a Weyl
invariant action we should introduce into \p{2.1} (or \p{2.1a}) an
auxiliary field $W(\xi)$ in such a way that the first and the second
term acquire the factor $W^p$ and $W^{p+1}$, respectively:
\begin{equation}\label{wa}
S_{{ D,p}} = (\a ')^{-{1 \over 2}}
\int d^{{ p+1}} \xi \hskip9pt e(\xi )
\left( - W^p e^{\mu}_{a} \partial_{\mu } x^{\underline m}
u^{a}_{{\underline m}} + pW^{p+1} \right).
\end{equation}
Then, for
example, \p{2.11} takes the form
\begin{equation}\label{w}
dx^{\underline m} = We_{a} u^{{\underline m}a},
\end{equation}
and \p{2.11} is obtained from \p{w} by gauge fixing $W=1$. We shall
encounter this situation when studying doubly supersymmetric p--branes.

On the other hand $W$ can be eliminated from \p{wa} by substituting
into \p{wa} the solution to its equation of motion
\begin{equation}\label{wi}
W={1\over{p+1}}\partial_{m} x^{\underline m}
u^{~a}_{\underline m}e_{~a}^{m}
\end{equation}
 This results in a
Weyl invariant p-brane action considered previously in \cite{weyl}.
Note that \p{wi} does not produce any new relations between the
variables, since it is just a consequence of \p{w}.

\section{Intrinsic and extrinsic geometry of the embedded surface
}

The geometrical approach \cite{lr,om} implies that the world
surface of a $p$--brane (and hence its equations of motion) is described
by the pullback of the Cartan forms $\Omega^{\underline{ab}}(d)$.

To do this we should replace Eqs.\p{2.11}, \p{2.13} by some equivalent
system of equations on the differential forms \p{2.5} and the world
surface vielbein $e^{a}$.

Eq. \p{2.13} is almost of the required type. Indeed, projecting \p{2.13}
onto the moving frame vectors $u^{a}_{{\underline m}}$ and
$u^{{i}}_{{\underline m}}$ we get the following two equations
equivalent to \p{2.13}
\begin{equation}\label{2.33}
\partial _{m }
\left(ee^{m{a}}\right) =
e e^{{m}}_{{b}} \Omega^{{ba}}_{{m}},
\end{equation}

\begin{equation}\label{2.34}
e^{{m}}_{{a}} \Omega^{{ai}}_{{m}}= 0,
\end{equation}

and eq. \p{2.11} can be replaced by its integrability conditions
\begin{equation}\label{2.35}
0 = ddx^{{\underline m}} =
d \left( e^{a}u_{a}^{\underline m} \right),
\end{equation}
which ensures the possibility of finding $x^{\underline{m}}$ if
$e_{am}$ and $u^{a\underline m}$ are derived from \p{2.12}, \p{2.13} with
the Cartan forms satisfying the Maurer--Cartan equations
\p{2.42}--\p{2.44}.

Projecting \p{2.35} onto $u^{{a}}_{{\underline m}}$ and
$u^{{i}}_{{\underline m}}$ we get the {\sl metricity condition} for
the induced connection on the world surface  defined as the
pullback of the Cartan form
$\Omega^{{ab}}= u^{{a}}du^{{b}}$:
\begin{equation}\label{2.36}
De^{a}\equiv T^{a} \equiv de^{a}
- e_{b} \Omega ^{ba} = 0
\end{equation}
(where $T^a_{bc}e^be^c$ is the intrinsic torsion of the surface),
and the
relation
\begin{equation}\label{2.37}
e_{a} \Omega^{{ai}} = 0
\end{equation}
{}From \p{2.36} it follows that
$\Omega ^{{ab}}$ is completely determined in terms
$e^{~a}_{m}$
\begin{eqnarray}\label{2.38}
e^{m}_{a} \Omega_{{ m bc }} \equiv
e^{m}_{a} u_{{\underline m b }} \partial_{{m}}
u^{{\underline m}}_{c}
%\equiv
%e^{m}_{a}
%\left(
%\Omega_{{ [\{f\} \{g\}]\{h\} }} -
%\Omega_{{ [\{g\} \{h\}]\{f\} }} +
%\Omega_{{ [\{h\} \{f\}]\{g\} }} \right) =
%\nn
= e^{m}_{a} e^{{n}}_{b} \partial_{{[m}} e_{{n] c}}
- e^{m}_{b} e^{{n}}_{{c}} \partial_{{[m}} e_{{n] a}}
+ e^{m}_{c} e^{{n}}_{{a}} \partial_{{[m}} e_{{n] b}}
\end{eqnarray}
and \p{2.43} is the induced Riemann curvature of the surface.

Eq. \p{2.37} ensures a symmetry property of the Riemann tensor \p{2.43}
(a Bianchi identity),
namely,
$~e^a R^{~b}_{a} = 0$, and identifies $e_{am} \Omega_n^{{ai}}$ with the
components $K^i_{mn}$ of the second differential form of the surface.
Indeed, by definition
\begin{equation}\label{2.39}
K^{i}_{{mn}} \equiv
{u}^{i}_{{\underline m}}
\partial_{m}
\partial_{n} x^{\underline m} ,
\end{equation}
where eq. \p{1.3} should be taken into account.
Substituting eq.\p{2.11} into \p{2.39} taking into account
\p{2.37} we get
\begin{equation}\label{2.40}
K^{i}_{{mn}} =
-e_{{n a}} \Omega^{{ai}}_{{m}} ,
\end{equation}

Now it is easy to see that eq.\p{2.34} means the vanishing of
the average extrinsic curvatures of the surface
and, hence, defines the embedding into the flat space--time
of a minimal surface
\begin{equation}\label{2.41}
K^{i}_{mn} g^{m n} \equiv h^{i}
= - e^{m}_{a} \Omega^{a~i}_{m}
= 0.
\end{equation}

To complete the identification of the Cartan forms with the
geometrical  characteristics of the surface note that
$\Omega^{~ij}_{m}$ coincides with the extrinsic torsion of the surface
in the target space.

The system of the Maurer--Cartan equations \p{2.42}--\p{2.44} supplemented
with eqs. \p{2.36}, \p{2.37} completely determines a surface
$x^{\underline{m}} = x^{\underline{m}} (\xi) $
up to its rotations and displacements in the target space. When eqs.
\p{2.42} -- \p{2.44} are rewritten as ones determining the second
quadratic form \p{2.39}, \p{2.40} and the extrinsic torsion
$\Omega^{~ij}_{m}$, they coincide with the Codazzi, Gauss and Ricci
equations, respectively.  Then eqs. \p{2.36}, \p{2.37}  are identically
satisfied.

In addition to the Codazzi, Gauss and Ricci equations the classical motion
of the bosonic $p$--brane is characterized by Eq. \p{2.41} which means
that the world surface is a minimal surface. This completes the description
of the bosonic $p$--brane theory in terms of surface theory.

The geometrical approach can also be applied for studying p--branes in a
curved target space, then one should use the Cartan equations
for the forms in the curved manifold, which, in general, involves
its torsion and curvature. Flat superspace is one of the
examples of this more general situation.

 \chapter{Towards a doubly supersymmetric geometrical approach to
super--$p$--branes}

To develop the geometrical approach in application to super--$p$--branes
we should determine the notion of the local moving frame in a
flat superspace parameterized by bosonic vector coordinate
$x^{\underline{m}}$ and fermionic spinor coordinate
$\theta^{\underline{\mu}}$
\begin{equation}\label{3.0}
z^{\underline{M}} = (x^{\underline{m}}, \theta^{\underline{\mu}} ).
\qquad
\end{equation}
As we will see, this naturally leads to spinor Lorentz harmonics
as the fundamental constituents of the moving frame.

\section{Spinor Lorentz harmonics as a moving frame in superspace}
Let us consider a supersymmetric
basis
\begin{equation}\label{3.1}
\pi^{\underline{m}} \equiv dx^{\underline{m}}
- i d\theta \Gamma^{\underline{m}} \theta ,\qquad
d\theta^{\underline{\mu}}
\end{equation}
in the space cotangent to the flat superspace.

Because the structure group of the flat superspace (as well as of the
curved one \cite{supergrav}) is  the double covering group
{\it Spin}$(1,D-1)$ of the Lorentz group $SO(1,D-1)$,
an arbitrary local frame in the flat
superspace can be obtained from \p{3.1} by $SO(1,D-1)$ rotations, i.e.
\begin{equation}\label{3.3}
\pi^{\underline{a}}
\equiv \pi^{\underline{m}} u^{\underline{a}}_{\underline{m}}
\qquad
\psi^{\underline{\a}} \equiv
d\theta^{\underline{\mu}}
v^{~\underline{\a}}_{\underline{\mu}}
\qquad
\end{equation}

The vector part of \p{3.1} is transformed by a matrix
$||u^{\underline{a}}_{\underline{m}}||$
from the vector representation
of the $SO(D-1)$, and the spinor part is transformed by a
matrix
$||v^{~\underline{\a}}_{\underline{\mu}} ||$
from the spinor
representation of $SO(1,D-1)$, or, strictly speaking,
\begin{equation}\label{3.4}
||v^{~\underline{\a}}_{\underline{\mu}} ||
{}~~~\in~~~ \hbox{ {\it Spin}}(1,D-1).
\qquad
\end{equation}

Thus, the components
of the new moving frame in the basis \p{3.1} are
$u^{\underline{a}}_{\underline{m}}$ and
$v^{~\underline{\a}}_{\underline{\mu}}$,
but since the transformations of
the vector and spinor sector are characterized by the same parameters,
$u^{\underline{a}}_{\underline{m}}$ and
$v^{~\underline{\a}}_{\underline{\mu}}$ are connected
by the relation expressing the vector representation
through the fundamental spinor representation of $Spin(1,D-1)$:
\begin{eqnarray}\label{3.7}
u^{\underline{a}}_{\underline{m}}
&\equiv& {{ 1}\over { 2^{\nu}}} v^{\underline{\a}}_{\underline
{\mu} } (\Gamma_{\underline{m}})^{\underline{\mu} \underline{\nu}}
v^{\underline{\b}}_{\underline{\nu}} (\Gamma^{\underline{a}})
_{\underline{\a}\underline{\b}}\nn
&\equiv & {{ 1}\over { 2^{\nu}}} v^{\underline{\mu}}_{{\underline
\g }} (\Gamma^{{\underline a}})^{\underline{\g}\underline{\d}}
v^{\underline{\nu}}_{\underline{\d}}(\Gamma_{\underline{m}})_{
\underline{\mu}\underline{\nu}}
\end{eqnarray}
where
\begin{equation}\label{3.9}
v^{-1} \equiv ||v_{\underline{\a}}^{~\underline{\mu}} ||
{}~~~\in~~~ \hbox{ {\it Spin}}(1,D-1)
\end{equation}
is the matrix inverse to \p{3.4}.

Hence, in superspace, the vector components of the local moving frame are
naturally composed of the bosonic spinor components, the latter playing
the basic role in the doubly supersymmetric  \cite{stv}--\cite{bers94}
as well as
twistor--like Lorentz--harmonic
approach \cite{bh,ghs,gds,ght,bzst,bzst1,bzm} (see also
 \cite{balach,wieg})\footnote{
The vector moving frame variables
$ u^{\underline{a}}_{\underline{m}}$ are just
vector Lorentz harmonics introduced by Sokatchev \cite{sok} as
an extension of the concept of harmonic variables \cite{gikos} to
noncompact groups of space--time symmetry.
For the first time
the vector moving frame
composed of a spinor one  was introduced by Newman
and Penrose \cite{npen} in application to General
Relativity. In application to superparticles and superstrings vector
harmonics, part of which was composed of twistor--like variables, were
considered in \cite{nis} (this approach was further developed in
\cite{kr}). Wiegmann \cite{wieg} used the composed moving frame for
the calculation of the anomalies in spinning and heterotic string.}.

Below in this section we will present some basic properties of the spinor
moving frame \cite{bh,bz0,ghs,gds,bzst,bzst1,bzm,ght,bzm1} required for
further consideration (see Appendix {\bf A} for detailes).

Due to \p{3.9}, Eq. \p{3.7} can be rewritten as follows
\begin{equation}\label{3.10}
u^{\underline{a}}_{\underline{m}}
(\Gamma_{\underline{a}})_{\underline{\a}\underline{\b}}
\equiv
v^{\underline{\mu}}_{\underline{\a}}
(\Gamma_{{\underline m}})_{\underline{\mu}\underline{\nu}}
v^{\underline{\nu}}_{\underline{\b}},
\end{equation}
which reflects the transformation properties of the
$\Gamma$--matrices with respect to the Lorentz group.

As in the bosonic case, for further description of the  embedding of a
super--$p$--brane world surface into the flat target superspace, the
$SO(1,p) \times SO(D-p-1)$ invariant splitting of the composed vector
moving frame $
u^{~\underline{a}}_{\underline{m}} = ( u^{~a}_{\underline{m}}~,~
u^{~i}_{\underline{m}})$ is required. As a consequence of \p{3.10}
this splitting is obtained by choosing an $SO(1,p) \times
SO(D-p-1)$ invariant representation for the $\Gamma$--matrices
\begin{equation}\label{3.13}
\Gamma^{\underline{a}}
= ( \Gamma^{a}, \Gamma^{i} )
\end{equation}
with block--diagonal $\Gamma^{a}$ and anti--diagonal $\Gamma^{i}$
(for $D=11,~p=2$ and $D=10, ~p=1$ such a representation is presented in
Appendix A), and decomposing the
harmonic matrix into two rectangular blocks
\begin{equation}\label{3.12}
v^{~\underline{\mu}}_{\underline{\a}} =
( v^{~~\underline{\mu}}_{\a q}~,~v^{\a\underline{\mu}}_{\dot q} ) ,
\end{equation}
where the index $\a$ corresponds to the spinor representation of $SO(1,p)$
and $(p,~\dot q)$ stand for two (in general non--equivalent) spinor
representations of $SO(D-p-1)$.
Then, the relation \p{3.10}
 splits into three $SO(1,p)\times SO(D-p-1)$ invariant
relations
\begin{equation}\label{3.15}
\d_{qp} (\gamma_a )_{\a \b}
u^{~a}_{\underline{ m}} =
v_{\a q}
%( \Gamma_{\underline{ m}} )_{\underline{\mu } \underline{\nu}}
%v^{\underline{\nu}}_{\b p} ,
%\equiv v_{\a q}
\Gamma_{\underline{ m}}
v_{\b p} ,
\end{equation}
\begin{equation}\label{3.16}
\d_{\dot{q} \dot{p}} (\gamma_{a})^{\a \b}
u^{a}_{\underline{ m}} =
v^{\a }_{\dot q}
\Gamma_{\underline{m}}
v^{\b }_{\dot p} ,
\end{equation}
\begin{equation}\label{3.17}
\d^{\a}_{\b} \gamma ^{i}_{q \dot p}
u^{i}_{\underline{ m}}
= v_{\a q}
\Gamma_{\underline{ m}}
v^{\b }_{\dot p} ,
\end{equation}
where $\gamma_{a}^{\a \b}$ and $\gamma ^{i}_{q \dot p}$ are the $SO(1,p)$
and $SO(D-p-1)$ $\g$--matrices, respectively (see Appendix {\bf A} for the
D=11, p=2 case).

We see that in \p{3.15}--\p{3.17} $(v^{\underline{\mu}}_{\a q},
v^{\a\underline{\mu}}_{\dot q})$, as well as $(u^{a}_{\underline{
m}},u^{i}_{\underline{ m}})$, are determined up to the local
$SO(1,p)\times SO(D-p-1)$ transformations and can be identified with
spinor Lorentz harmonics parametrizing the coset space
${SO(1,D-1)}\over{SO(1,p)\times SO(D-p-1)}$. Note that in contrast to the
splitting of vectors that of the spinors results in
multiplicative structure of the
${SO(1,p)\times SO(D-p-1)}$ spinor indices.

This basic notion on the local moving frame
in superspace is sufficient for developing the geometrical approach to
super--$p$--branes.

In the conventional formulation of super--$p$--branes
\cite{m0,m1,t1,m1d,m} one considers the embedding of a bosonic world
surface spanned by  a $p$--brane moving in a target superspace. In the
approach under consideration this embedding can be described by an action
which is the sum of the conventional Wess--Zumino term
\cite{gschw84,m0,m1,t1} plus an analog of Eq. \p{2.1} where
$\partial_{m} x^{\underline m} (\xi)$ is replaced with  the
$\Pi^{~\underline{m}}_{n}$
component of the supersymmetric Cartan form
$\Pi^{~\underline{m}}= dx^{\underline m} - i d\theta
\Gamma^{\underline{m}} \theta$ and $u^{a}_{\underline{m}}(\xi)$ is composed
of $v^{~\underline{\mu}}_{\underline{\a}} (\xi)$ (Eq.\p{3.15})
\cite{bzst,bzst1,bzm,bzm1}.  Then, to derive equations of motion of
$v^{~\underline{\mu}}_{\underline{\a}} (\xi)$,
one should, as in the bosonic case (eqs. \p{2.7}), consider the variations
of $v^{~\underline{\mu}}_{\underline{\a}} (\xi^m)$, which do not
violate the condition \p{3.4}, \p{3.7}.  These variations are:
$$
\delta v^{~{\underline \mu}}_{{\underline
\a}} = -O_{\underline \a}^{~\underline
\b}v_{\underline{\b}}^{\underline{\mu}}= - {1\over 4} O
^{{\underline a}{\underline b}}(\Gamma_{\underline{a}
\underline{b}})^{~\underline \g}_{\underline \a}
v^{~{\underline\mu}}_{{\underline\g}} .
$$
And
\begin{equation}\label{3.17.}
dv^{~{\underline \mu}}_{{\underline
\a}} = - {1\over 4} \Om
^{{\underline a}{\underline b}}(d)(\Gamma_{\underline{a}
\underline{b}})^{~\underline \g}_{\underline \a}
v^{~{\underline\mu}}_{{\underline\g}} .
\end{equation}
(compare with \p{2.7}, \p{2.5}).

The studying of the constraints and the equations of motion of
super--$p$--branes in the Lorentz harmonic formulation was performed in
\cite{bzst,bzst1,bzm,bzm1}, so we only note that this formulation can be
regarded as a component version of a doubly supersymmetric $p$--brane
model \cite{stv}--\cite{bers94}, and proceed with developing the
geometrical approach to the latter.

\section{Geometrodynamical condition, twistor constraint and geometrical
framework for the description of super--$p$--branes}

In the doubly supersymmetric formulation of
super--$p$--branes \cite{stv}--\cite{bers94}, their dynamics is described by
embedding the world supersurface
$$
z^M \equiv ( \xi^m , \eta^{\a q} ),
$$
into the target superspace \p{3.0}, which is further considered to be flat,
$$
z^{\underline{M}} = Z^{\underline{M}}(z^{M}).
$$
Note that supersurfaces under consideration the number of the
Grassmann directions is half of the
number of the target superspace
Grassmann directions.
An intrinsic world supersurface geometry is assumed to be characterized by
torsion constraints \cite{tp93,bers94}
\begin{eqnarray}\label{4.1}
T^{a} \equiv D E^{a} \equiv d E^{a} - E^{b}
w^{~a}_{b} &=& - i E^{\a q} E^{\b}_{q} \g^{a}_{\a \b} , \nn
T^{\a q} \equiv D E^{\a q}& \equiv & d E^{\a q} - E^{\b q} w^{~\a}_{\b} +
E^{\a p} A^{~q}_{p}\nn
 &&= -  E^{a } E^{\b p} T^{~~\a q}_{a\b p} - E^{a} E^{b}
T^{~~\a q}_{ab},
\end{eqnarray}
where
\begin{equation}\label{4.2}
E^{a} \equiv d z^{M} E^{~a}_{M} \qquad ~~and~~ \qquad
E^{\a q} \equiv d z^{M} E^{~\a q}_{M}
\end{equation}
are the vector and spinor world supersurface vielbein forms, and
$w^{~a}_{b}$, $A^{~q}_{p}$ are the components of an $SO(1,p)\times
SO(D-p-1)$ connection.

We should stress that the only essential torsion constraint in
\p{4.1} is $T^{~~~~~a}_{\a q~\b p}~=~ -2i \d_{qp} \g^{a}_{\a\b}$ ensuring
flat supersurface limit. The other torsion constraints are obtained by
solving for the Bianchi identities and
redefining vielbeins and connections.

Eqs. \p{4.1} imply, in particular, the following
anticommutation relations for supercovariant spinor derivatives
\begin{equation}\label{4.3}
\{ D_{\a q}~,~ D_{\b p} \} = 2i \d_{qp} \g^{a}_{~\a \b} D_{a} +R_{\a q~
\b p},
\end{equation}
where $R_{\a q~\b p}$ are components of intrinsic $SO(1,p)\times
SO(D-p-1)$ curvature
\begin{equation}\label{4.19}
SO(1,p):\qquad R^{ab}
= dw^{ab} - w^{ac} w^{~b}_{c},\qquad
R^{\a\b}\sim R^{ab} \g_{ab}^{\a\b},
\end{equation}
\begin{equation}\label{4.20}
SO(D-p-1):\qquad R^{ij}
= dA^{ij} + A^{ik} A^{kj},\qquad
R^{pq} \sim R^{ij} \g_{ij}^{~~pq} .
\end{equation}

Note that at least for superstrings (p=1) and supermembranes (p=2)
the constraints \p{4.1} are intact under super--Weyl transformations
of the supervielbeins
\begin{equation}\label{weyl}
\hat E^a=W^2E^a,\qquad
\hat E^{\a q}=WE^{\a q}-iE^b\g_b^{\a\b}D^q_\b W
\end{equation}
and corresponding transformations of the connection forms. This property
will be used below for studying a D=11 supermembrane  and D=10
superstrings.

Let us consider the pullback of the one--forms
\p{3.1} onto the world supersurface \begin{equation}\label{4.4}
\Pi^{\underline{m}} \equiv d X^{\underline{m}} - i d \Theta
\Gamma^{\underline{m}} \Theta = E^{\a q} \Pi^{\underline{m}}_{\a q} +
E^{a} \Pi^{\underline{m}} _{a}, \end{equation} \begin{equation}\label{4.5}
d \Theta^{\underline{\mu}}
= E^{\a q} D_{\a q}\Theta^{\underline{\a}}+
E^{a} D_{a} \Theta^{\underline{\mu}}.
\end{equation}
where, in view of \p{4.3},
\begin{equation}\label{4.10}
2 i \d_{qp} \g^a_{\a \b}  D_a\Th^{\underline{\mu}} =
D_{\a q} D_{\b p}\Th^{\underline{\mu}} +
D_{\b p} D_{\a q}\Th^{\underline{\mu}} ,
\end{equation}

The twistor--like bosonic superfield
$ D_{\a q}\Theta^{\underline{\mu}}$
%and fermionic superfield
%$ \Psi^{\underline{\mu}}_{a} \equiv D_{a} \Theta^{\underline{\mu}}~$
plays
the basic role in the development of the doubly supersymmetric
geometrical approach.

The embedding of a super--$p$--brane world supersurface is specified by
a geometrodynamical condition
 \cite{stv}--\cite{bers94}, which requires the vanishing
of $\Pi^{\underline{m}}$  along
 the Grassmann  world supersurface directions:
\begin{equation}\label{4.6}
\Pi^{\underline{m}}_{\a q} \equiv
D_{\a q} X^{\underline{m}} - i D_{\a q} \Theta \Gamma^{\underline{m}} \Theta
= 0.
\end{equation}

The integrability condition for \p{4.6}
\begin{equation}\label{4.9} \d_{qp} \g^a_{~\a \b}
\Pi^{~\underline{m}}_{a} = D_{\a q}\Th
\Gamma^{\underline{m}}D_{\b p}\Th
\end{equation}
is called ``twistor constraint''.

Eq.\p{4.9} looks very much like Eq. \p{3.15} which relates the
spinor and vector Lorentz harmonics. So we argue that
$D_{\a q}\Th^{\underline{\mu}}$
can be identified (up to a scalar superfield factor
$W$)
with
$v^{~\underline{\mu}}_{\a q}$
\p{3.7}, \p{3.10}, and
$\Pi_{~a}^{\underline{m}}$,
can be identified with
$u_{~a}^{\underline{m}}$ \p{3.15},
(up to the square $W^{2}$ of the same factor):
\begin{equation}\label{4.11}
D_{\a q}
\Theta^{\underline{\mu}} =W
v^{~\underline{\mu}}_{\a q} ,
\end{equation}
\begin{equation}\label{4.12}
\Pi_{~a}^{\underline{m}}=
D_{a} X^{m} - D_{a} \Theta \Gamma^{\underline m} \Theta
= W^{2} u_{~a}^{\underline{m}} ,
\end{equation}
For a $D=11$, N=1 supermembrane a direct proof of
\p{4.11}, \p{4.12} is presented in the Appendix {\bf B}.

Thus, the spinor moving frame ($\equiv$ Lorentz harmonics),
which is the generalization of the Cartan moving frame to the case
of superspace, naturally appears  in the
doubly supersymmetric $p$--brane formulation.

Eq. \p{4.12}\footnote{
The leading component of this equation appears as an equation of motion in
the twistor--like Lorentz harmonic super--$p$--brane formulation
\cite{bzm,bzm1}.} can be regarded as the supersymmetric counterpart of
eq. \p{1.4}, and eq. \p{4.11} is a ``square root'' of  \p{4.12}.

In view of \p{4.6}, \p{4.10}, \p{4.11} and \p{4.12}
the one-forms \p{4.4}, \p{4.5} are expressed as follows
\begin{equation}\label{4.13}
\Pi^{\underline{m}} =
E^{a} W^2 u^{\underline{m}}_{a} ,
\end{equation}
\begin{equation}\label{4.14}
d\Th^{\underline{\mu}} =W
E^{\a q}
v^{~~\underline{\mu}}_{\a q} +
E^{a}
D_{a}\Th^{\underline{\mu}}.
\end{equation}
{}From \p{4.13}, \p{4.14} we conclude that an
induced metric $\Pi^{\underline{m}}\Pi_{\underline{m}}$ on the
supersurface coincides with an intrinsic metric $E^aE_a$ up to the scale
factor ${W^4}$. Thus $W$ plays the role of a rescaling factor of the
intrinsic metric.

To study the properties of the supersurface embedding it is necessary to
consider the integrability conditions for eqs. \p{4.13}, \p{4.14}
with taking into account \p{4.6}
\begin{equation}\label{4.7}
d \Pi^{\underline{m}}
= - i d \Theta \Gamma^{\underline{m}} d \Theta ,
\end{equation}
\begin{equation}\label{4.8}
dd \Theta^{\underline{\mu}}= 0 .
\end{equation}
Eqs. \p{4.7}, \p{4.8} are the pullback of
the Maurer--Cartan equations for supertranslations in the flat superspace
\cite{supergrav} (they should not be confused with the Maurer--Cartan
equations \p{2.42} -- \p{2.44} for $SO(1,D-1)$).

Substituting into \p{4.7}, \p{4.8} the expression for $\Pi^{\underline{m}}$
and $d \Theta^{\underline{\mu}}$ in terms of the harmonics (\p{4.13},
\p{4.14}) and projecting onto the $SO(1,p)\times SO(d-p-1)$ directions we
get
\begin{equation}\label{4.7a}
d(W^2E^a)-(W^2E^b)\Om_b^{~a}= -
id\Theta\Gamma^{\underline{m}}d\Theta u_{\underline{m}}^a\equiv
T_{\rm ind}^a,
\end{equation}
\begin{equation}\label{4.7b}
(W^2E^b)\Om_{b}^{~i}=
id\Theta\Gamma^{\underline{m}}d\Theta u_{\underline{m}}^i,
\end{equation}
\begin{equation}\label{4.8a}
d(WE^{\a p})-(WE^{\b q})\Om_{\b q}^{~\a p}=T_{\rm ind}^{\a
p},
\end{equation}
\begin{equation}\label{4.8b}
(WE^{\b q})\Om_{\b q}^{~\a
\dot p}=T_{\rm ind}^{\a\dot p}.
\end{equation}
Eqs. \p{4.7a}, \p{4.8a} (see \p{3.17.} for the $\Om$--notation)
define the components $\Om^{ab},~\Om_{\a p}^{~\b q}=\Om_\a^\b\d_q^p-
\Om_{p}^{~q}\d_\a^\b$ of
induced connection on the supersurface, with $T_{\rm ind}^a,$ $T_{\rm
ind}^{\a p}$ being the components of induced torsion. (Since the explicit
form of $T_{\rm ind}^{\a p},~T_{\rm ind}^{\a \dot p}$ is rather
complicated, we do not present it here). The $\Om$--forms in
\p{4.7a}--\p{4.8b} satisfy the Maurer--Cartan equations
\p{2.42}--\p{2.44}, and eqs. \p{4.7b}, \p{4.8b} ensure symmetry properties
of the Riemann tensor in the presence of induced torsion.

One can see that, in general, the supersurface geometry induced
by the embedding under consideration differs from the intrinsic
geometry defined by eqs. \p{4.1}, and the set of equations \p{4.6},
\p{4.11}, \p{4.12}, \p{4.7a}--\p{4.8b} and the Maurer--Cartan equations
\p{2.42}--\p{2.44}
relates the two kinds of geometry.  In particular, from \p{4.7a}, \p{4.1}
it follows that the spinor-spinor components of $W^{-2}T_{ind}^a$ coincide
with that of $T^a$.

To get all the consequences of eqs. \p{4.7}, \p{4.8} (or
\p{4.7a}--\p{4.8b}), and eqs. \p{2.42}--\p{2.44} one has to solve for
their
components in the basis $$ E^{\a q} E^{\b p}~,
\qquad E^{a} E^{\b p}~, \qquad E^{a} E^{b}~. $$

It turns out, however, that for a supersurface with
$n > 1$  only equations corresponding to the spinor--spinor components
are independent (see Appendix {\bf C})\footnote{
Of course, the choice of the independent relations is not unique.}.
This means that all other consequences
can be derived
by taking the spinor derivatives  of the spinor-spinor components and
using \p{4.3}. For example, the independent consistency equations contained in
\p{4.7}, \p{4.8} are \p{4.10}, \p{4.9}.

To compare the induced geometry with the intrinsic one it is convenient to
introduce covariant objects reflecting the difference between
induced and intrinsic connection:
\begin{equation}\label{4.29}
%\Omega^{~ab}_{\a q} \equiv
\Omega^{ab}(D) =\Omega^{ab} (d)- w^{ab},
\end{equation}
\begin{equation}\label{4.30}
%\Omega^{ij}_{\a q} \equiv
\Omega^{ij}(D) =
\Omega^{ij} (d)
- A^{ij},
\end{equation}
\begin{equation}\label{4.28}
\Omega^{ai}(D)=
\Omega^{ai} (d).
\end{equation}
Note that in the bosonic case (Chapter 1), where the induced and
intrinsic geometry coincide, $\Omega^{~ab}(D)=0$.

In terms of \p{4.29}--\p{4.28} the Maurer--Cartan equations take the form
\begin{equation}\label{4.16}
D^{ind}\Om^{ai}=D \Omega ^{ai} -
\Omega^{a}_{~b} (D) \Omega^{bi} +
\Omega^{aj} \Omega^{ji} (D) = 0,
\end{equation}
\begin{equation}\label{4.17}
R_{ind}^{ab}=D \Omega^{ab} (D)
-\Omega^{a}_{~c} (D) \Omega^{cb} (D)+R^{ab}
 =
\Omega^{ai} \Omega^{bi},
\end{equation}
\begin{equation}\label{4.18}
R_{ind}^{ij}=D \Omega ^{ij} (D) +
\Omega ^{ij^\prime} (D) \Omega^{j^\prime j} (D)+R^{ij}=
\Omega ^{ai} \Omega^{~j}_{a}.
\end{equation}
The independent (spinor--spinor) components of \p{4.16}--\p{4.18} are
\begin{eqnarray}\label{4.25}
-2i \d_{qp} \g^{b}_{\a \b} \Omega^{~ai}_{b}~=&
D_{\a q} \Omega^{~ai}_{\b p} +
\Omega^{~~a}_{\a q~b} \Omega^{~~bi}_{\b p}
- \Omega^{~~aj}_{\a q} \Omega ^{~ji}_{\b p}
\nn
&+((\a q) \leftrightarrow (\b p)),
\end{eqnarray}
\begin{eqnarray}\label{4.26}
-2i \d_{qp} \g^{c}_{\a \b} \Omega^{~ab}_{c}~=&
D_{\a q} \Omega^{~ab}_{\b p} +
\Omega^{~~a}_{\a q~c} \Omega^{~~cb }_{\b p}
+ \Omega^{~~aj}_{\a q} \Omega ^{~b j}_{\b p}+R_{\a q\b p}^{ab}
\nn
&+( (\a q) \leftrightarrow (\b p)) ,
\end{eqnarray}
\begin{eqnarray}\label{4.27}
-2i \d_{qp} \g^{c}_{\a \b} \Omega^{~ij}_{c}~ =&
D_{\a q} \Omega^{~ij}_{\b p} -
\Omega^{~~ik}_{\a q} \Omega^{~~kj }_{\b p}
- \Omega^{~~ai}_{\a q} \Omega ^{~~~j}_{\b p a } +R_{\a q\b p}^{ij}
\nn
&+ ((\a q) \leftrightarrow (\b p) ),
\end{eqnarray}

Eqs.\p{4.16}--\p{4.18} (or \p{4.25} -- \p{4.27}), with taking into account
\p{4.7a}--\p{4.8b}, can be regarded as supersymmetric analogs of Codazzi,
Gauss and Ricci equations.

\section{Minimal supersurface embedding into flat superspace}

We have seen in the bosonic case that the equations of motion of a
$p$--brane determine minimal embedding of the world surface, and the
minimal surface is characterized by the traceless second fundamental form
(eqs.\p{2.39}--\p{2.41}).

In the doubly supersymmetric case we shall also assume that the equations
of motion of a super--$p$--brane determine a minimal embedding of the
world supersurface (and vice versa), which is characterized by the
vanishing trace of a supersymmetric counterpart of the bosonic second
fundamental form.

An appropriate $SO(1,p)\times SO(D-p-1)$--valued bilinear form on the
supersurface has the following components
\begin{equation}\label{k1}
K^i_{AB}=(K^i_{ab}, K^i_{\a pb},K^i_{\a p\b q}),
\end{equation}
\begin{equation}\label{k2}
K^{\g \dot r}_{AB}=(K^{\g \dot r}_{ab},
K^{\g \dot r}_{\a p b},K^{\g \dot r}_{\a p\b q}),
\end{equation}
and is symmetric with respect to the permutations of the vector--vector
and vector--spinor indices and antisymmetric with
respect to the permutations of the pairs of spinor indices ($\a p,\b q$).

Since the structure group in the supersurface tangent space is
$SO(1,p)\times SO(D-p-1)$, each component in \p{k1}, \p{k2} transforms
independently and, thus, can be regarded as an independent supersymmetric
bilinear form. For describing the embedding in question we suppose that it
is sufficient to determine the $K^i_{ab}$ and $K^{\g \dot r}_{\a p\b q}$
component of \p{k1}, \p{k2}  as supersymmetric analogs of the second
fundamental form \p{2.39}, \p{2.40}, and take
\begin{equation}\label{k11}
K^i_{ab}=2\left(D_{\{a}\Pi^{\underline{m}}_{b\}}\right)u^i_{{\underline
m}}=-2W^2\Om_{\{ab\}}^{~~i}, \end{equation}
\begin{equation}\label{k21}
K^{\g \dot r}_{\a p\b q}=2\left(D_{[\a p}D_{\b
q]}\Th^{\underline\mu}\right)v_{\underline\mu}^{\g\dot r}=-{W}\Om_{\a
q}^{ai}(\g_{a})_{\b}^{\g}(\g_i)_q^{\dot r}-((\a p)\leftrightarrow (\b q)).
\end{equation}
As in the bosonic case, \p{k11}, \p{k21} are expressed through the
components of $\Om^{ai}(d)$.

We assume that the minimal embedding of the supersurface into the flat
superspace is characterized by \p{k11}, \p{k21} with vanishing traces:
\begin{equation}\label{k12}
K^{ai}_{a}=0=\Om_{a}^{ai}=i{1\over{p2^{[{p\over 2}]}}}\left(D_{\a
q}\left((\g_a)^{\a\b} \Om_{\b q}^{ai}\right)+\Om^a_{\a q
b}(\g_a)^{\a\b}\Om_{\b q}^{bi}-(\g_a)^{\a\b}\Om_{\a q}^{aj}\Om_{\b
q}^{ji}\right),
\end{equation}
\begin{equation}\label{k22}
K^{\a p,\g \dot r}_{\a p}=0=(\g_a)^{\g\a}\Om_{\a p}^{ai}(\g_i)^{p\dot r}
\end{equation}
where the r.h.s. of \p{k12}  follows from eq. \p{4.25}.

The sufficient condition for \p{k22} to hold is
\begin{equation}\label{4.38}
\g^{a ~\b}_{~\a} \Omega^{~~ai}_{\b q}  = 0 .
\end{equation}
And it is just this condition which follows from equations of motion of
the super--p--brane.

Indeed, the appropriate form of the equations of motion of
$\Theta^{\underline{\mu}} (z^M)$ can be obtained by
a superfield
generalization of the equation of motion of the field
$\theta^{\underline{\mu}}(\xi^m)$
arisen in the Lorentz harmonic super--$p$--brane formulation
\cite{bzm,bzm1}:
\begin{equation}\label{4.33}
\g^{a}_{\a\b} D_{a}\Theta^{\underline{\mu}}
v_{\underline{\mu}}^{\b\dot q} = 0.
\end{equation}
(Recall that
$v^{~~\underline{\mu}}_{\a q}
v_{\underline{\mu}}^{\g\dot r} = 0$).

Using eqs.~\p{4.10}, \p{4.11}, \p{3.17.} one can represent
the l.h.s. of \p{4.33} as
\begin{equation}\label{4.37}
2 i \d_{qp} (\g^a)_{\a\b}D_{a}\Theta^{\underline{\mu}}
v_{\underline{\mu}}^{\b\dot r}=
(\g_{a})_\a^\b\Omega^{ai}_{\b q}(\g_i)_{p}^{\dot r},
\end{equation}
from which we conclude that eq.~\p{4.33} holds when  $\Omega^{ai}_{\b q}$
satisfies eq.~\p{4.38} and vice versa. In view of \p{k11}, \p{k12} this
also leads to the equations of motion of $X^{{\underline m}}$ provided
$\Om^{ab}_{\a p}$ is restricted by consistency conditions to be
\begin{equation}\label{cor}
\Om^{ab}_{\a p}\g_a^{\a\b}=\varphi_{\a p}\g^{b\a\b},
\end{equation}
where $\varphi_{\a p}$ is a spinorial superfield (we shall encounter
this situation below).

Note that \p{k12} specifies the $X^{\underline m}$ components
along the directions orthogonal to the world supersurface. The
$X^{\underline m}$ components along the directions tangent to
the supersurface can be eliminated by fixing a gauge with respect to the
local symmetries.

In the next two chapters we shall consider in more detail some particular
features of the world supersurface embedding in the case of D=11
supermembranes and D=10 superstrings. For instance, eq. \p{cor} and
the equation of
motion \p{4.33} of a D=11, N=1 supermembrane and D=10, N=II superstrings
will appear as a consequence of the geometrodynamical condition \p{4.6}.

\chapter{N=1 supermembrane in D=11}

In the previous chapter we have obtained the system of equations
determining an embedding of a world supersurface into the flat target
superspace and relating intrinsic and induced geometry on the
supersurface.

Below we shall study these equations in application to a supermembrane (i.e.
$p=2$) in N=1, D=11 target superspace possessing $n=D-p-1=8$ world sheet
supersymmetries \cite{tp93}. In particular, we will see that the equation of
motion \p{4.33} is among the consequences of the geometrodynamical
condition \p{4.6}, and that the difference between the spinor components
of the intrinsic and
induced $SO(1,2)\times SO(8)$ connection of the world surface (eqs.~\p{4.29},
\p{4.30}) is due to the presence of the scale factor $W(\xi,\eta)$.

To show this let us consider one of the independent equations of
the integrability condition \p{4.8}, namely
\p{4.10}, and take into account \p{4.11} (the latter being the
consequence of the twistor constraint \p{4.9} and, hence, of the
geometrodynamical condition \p{4.6}):
\begin{equation}\label{5.1}
2i\d_{qp}\g^a_{\a\b}D_a\Th^{\underline\mu}=
(D_{\a q}W)v^{\underline\mu}_{\b p}+WD_{\a q}v^{\underline\mu}_{\b
p}+((\a q)\leftrightarrow (\b p)).
\end{equation}

For further consideration one should make use of the relation \p{3.17}
and that of Appendix A allowing one to express the covariant
differential of $v^{\underline\mu}_{\a q}$ in terms of
$\Om^{\underline{ab}}(D)$
(eqs. \p{4.29}--\p{4.28}):
\begin{equation}\label{5.2}
Dv^{~\underline{\mu}}_{\a q} =
{i \over 4} \e_{abc} \Omega^{ab}(D)
\g^{c~\b}_{~\a} v^{~~\underline{\mu}}_{\b q}
+ {1 \over 4} \Omega^{ij}(D) \g^{ij}_{~qp} v^{~~\underline{\mu}}_{\a p}
- {1 \over 4} \Omega^{ai}(D)
\g_{a~\a \b} \g^i_{q \dot p} v^{\b \underline{\mu}}_{\dot p}.
\end{equation}

Projecting \p{5.1} onto the ``orthogonal'' directions
$v_{\underline{\mu}\g\dot r},v_{\underline{\mu}\g r}$ one gets
\begin{eqnarray}\label{5.5}
2i\d_{qp} \g^a_{~\a \b}(D_{a}\Th^{\underline\mu})v_{\underline\mu\g r}
&= D_{\a q}W\e_{\b\g} \d_{pr} +
{{iW}\over 4}  \e_{abc}
\Omega^{~ab}_{\a q} \g^{c}_{\b\g} \d_{p r} +
{ W\over 4}
\Omega^{~ij}_{\a q} \e_{\b\g} \g^{ij}_{p r}\nn
&+( (\a q) \leftrightarrow (\b p) ) ,
\end{eqnarray}
\begin{equation}\label{5.4}
2i \d_{qp} \g^a_{~\a \b}(D_{a}\Th^{\underline\mu})v_{\underline\mu\g\dot r}
= {1\over 4}W\Omega^{~ai}_{\a q} \g_{a~\b \g} \g^i_{p \dot r} +
( (\a q) \leftrightarrow (\b p) ) ,
\end{equation}
from which we shall derive eq. \p{4.33}, \p{cor} and which will allow us to
express $\Omega^{~ab}_{\a q}\equiv\Omega^{~ab}(D_{\a q})$,
$\Omega^{~ij}_{\a q}\equiv\Omega^{~ij}(D_{\a q})$ in terms of $D_{\a
q}W$.

Now recall that for the supermembrane under consideration the
constraints \p{4.1} are invariant under the super--Weyl transformations
\p{weyl} of the supervielbeins and the corresponding transformations
of the intrinsic $SO(1,2)\times SO(8)$ connection
\begin{eqnarray}\label{con}
\hat\omega^{ab}&=\om^{ab}+2W^{-1}(D^bWE^a-
D^aWE^b)+{1\over 2}W^{-2}\e^{abc}E_cD_{\a q}WD^{\a q}W\nn
&+2iW^{-1}
\e^{abc}\g_{c\a}^{~\b}E^{\a q}D_{\b q}W,\nn
&\hat A^{ij}=A^{ij}+ { 1 \over W}\g^{ij}_{qp}D_{\a p}W +...
\end{eqnarray}
(where dots denote insignificant terms). This allows one to put W=1
in \p{5.1}, \p{5.4}, \p{5.5} without violating the constraints.

\section{Relation between intrinsic and induced connection}

Let us consider eq. \p{5.5} with $W=1$.

First of all note that the SO(8) irreps {\bf 56} and {\bf 160} cannot be
contained in $\Omega^{~ij}_{\a q}\g^{ij}_{p r}$ (due to the structure of
the other terms in \p{5.5}), hence its general form is
%\begin{equation}\label{5.18}
%\Omega^{~ij}_{\a [q}\g^{ij}_{p r]} = 0
%\end{equation}
%and
\begin{equation}\label{5.19}
\Omega^{~ij}_{\a q}\g^{ij}_{p r}=
%{2\over 3}\left(\Omega^{~ij}_{\a
%\{q}\g^{ij}_{p\}r}-\Omega^{~ij}_{\a
%\{q}\g^{ij}_{r\}p}\right)
%=\Omega^{~ij}_{\a
%\{q}\g^{ij}_{p\}r}+{1\over 2}
%\Omega^{~ij}_{\a r}\g^{ij}_{pq}.
\phi_{\a[p}\d_{r]q}.
\end{equation}
Then, let us decompose the r.h.s and the l.h.s. of
\p{5.5} onto the SO(1,2)
irreducible parts as follows
\begin{equation}\label{5.20}
{\rm in~~l.h.s.}\qquad 2i\g^a_{~\a \b} D_a \Theta^{\underline{\mu}}
v_{\underline{\mu}~\g r}  \equiv
\psi_{\{ \a \b \g\}r} +
{2\over 3}\e_{\g \{ \a} \psi_{\b \} r}
\end{equation}
\begin{equation}\label{5.21}
{\rm in~~r.h.s.}\qquad {i\over 4}\e_{abc}
\Omega^{~~ab}_{\a q}
\g^c_{~\b \g} \equiv
\k_{\{ \a \b \g\}q} +{2\over 3}
\e_{\a \{ \b}\k_{\g \} q}.
\end{equation}
The part of \p{5.5} being completely
symmetric in $\{ \a \b \g \}$  has
the form
\begin{equation}\label{5.22}
\d_{qp} \psi_{\{ \a \b \g\}r} =
2\k_{\{ \a \b \g\}~\{ q} \d_{p \} r}.
\end{equation}
Putting $q=p \not= r$ for any $r$ we get from Eq. \p{5.22}
\begin{equation}\label{5.23}
\psi_{\{ \a \b \g\}r} = 0=
\k_{\{ \a \b \g\} q},
\end{equation}
and, hence,
\begin{equation}\label{5.24}
{i\over 4}\e_{abc}
\Omega^{~~ab}_{\a q}
\g^c_{~\b \g}={i\over 6}
\e_{abc}\e_{\a \{ \b}\g^{c\d}_{~\g\}}
\Omega^{~~ab}_{\d q}.
\end{equation}
Then, contracting eq. \p{5.5} with $\e^{\a\b}$ and using \p{5.19} we
obtain
\begin{equation}\label{5.25}
{1\over 4}\Omega^{~ij}_{\a q} \g^{ij}_{pr} ={1\over
4}\phi_{\a[p}\d_{r]q}=
 - {i\over 2}\e_{abc}\Omega^{~~ab}_{\d [p}
\g^{c\d}_{~\a} \d_{r]q}.
\end{equation}
Substituting eq. \p{5.25} back into eq. \p{5.5} and taking into account
relations found above upon some manipulation with indices we finally get
\begin{equation}\label{0}
\Omega^{~ab}(D_{\a q}) =\Omega^{~ij}(D_{\a q}) =
(D_a\Th^{\underline\mu})v_{{\underline\mu}\g r}=0.
\end{equation}
Thus the pull back of $d\Th^{\underline\mu}$ is expressed in terms of
the spinor Lorentz harmonics and
$\psi_{\{\a\b\g\}\dot q}$ as follows:
\begin{equation}\label{5.31}
d \Theta^{\underline{\mu}} = E^{\a q}v^{~~\underline{\mu}}_{\a q} -
E^a {1 \over 2} \g^{\a \b}_a
\psi_{\{ \a \b \g\}\dot q}
v^{\g \underline{\mu}}_{\dot q}.
\end{equation}
If one performs the inverse super--Weyl transformation the r.h.s. of
\p{0} will become nonzero:
\begin{equation}\label{5.30}
{i \over 8} W \e_{abc} \Omega^{~ab}(D_{\a q}) \g^{c}_{\b\g} =
 \e_{\a \{ \b }  D_{\g \} q}W
\qquad or \qquad \Omega^{~ab}(D_{\a q}) =
 2i\e^{abc} \g^{~~~\b}_{c~\a}
 {1\over W} D_{\b q}W,
\end{equation}
\begin{equation}\label{5.29}
\Omega^{~ij}(D_{\a q}) =
\g^{ij}_{qp}
 { 1 \over W} D_{\a p}W,
\end{equation}
and
\begin{equation}\label{5.29a}
i\g^a_{\a\b}(D_a\Th^{\underline\mu})v_{{\underline\mu}\g r}
=2D_{\{\a r}W\e_{\b\}\g}.
\end{equation}

We see that the difference between the $E^{\a q}$ components of the
induced and the intrinsic $SO(1,2)\times SO(8)$ connection (eqs.
\p{5.30}, \p{5.29}) is due to the superfield $W$, and $\Om^{ai}(D_{\a
q})$ is
restricted to be of the form \p{5.17}. Further on, eqs.
\p{4.25}--\p{4.27} allow one to determine the $E^a$ components of
$\Om^{\underline{ab}}(D)$ in terms of their spinor components.
Note that \p{5.30} is a particular case of \p{cor}.

 Because of the
super-Weyl invariance of the d=2+1, n=8 supergravity constraints,
the superfield $W$ can be gauged away of the theory, so that the
intrinsic and induced geometry on the world supersurface coincide
at least for the spinor components of the connections. Note that
then the vector component of $\Om^{ai}$ satisfies the condition similar
to \p{2.37} in the bosonic case
$$
E_aE^b\Om^{ai}_b=0
$$
(which follows from \p{4.7b}, \p{5.31})
and can be regarded as the vector part \p{k11} of the supersymmetric analog
of the second fundamental form.

The difference $\Om^{ab}(D_c)$ between the vector components of
intrinsic and induced spin connection is due to nonzero components
$T^a_{bc}$ of induced torsion \p{4.7a} while intrinsic $T^a_{bc}$ was
chosen to be zero \p{4.1}. To completely identify the two connections one
should redefine $\om^{ab}_c$ in such a way that $\Om^{ab}(D_c)=0$, of
course then $T^a_{bc}$
will become non--zero  in the constraints \p{4.1}.

\section{Minimal embedding
of the supermembrane world surface.}

To find the restrictions on $\Omega^{~ai}_{\a q}$ and
$D_a\Th^{\underline\mu}$ which follow from
eq.\p{5.4} we decompose the r.h.s. and the l.h.s. of \p{5.4}
onto the irreducible
representations of $SO(1,2)$:
\begin{equation}\label{5.6}
{\rm in~~ l.h.s.}\qquad 2i\g^a_{~\a \b} D_a \Theta^{\underline{\mu}}
v_{\underline{\mu}~\g \dot r}  \equiv
\psi_{\{ \a \b \g\}\dot r} +
{2 \over 3} \e_{\g \{ \a} \psi_{\b \} \dot r}
\end{equation}
\begin{equation}\label{5.7}
{\rm in~~ r.h.s.}\qquad{1\over 4}\g^a_{~\a \b}\Omega^{~~ai}_{\g q} \equiv
\kappa^{i}_{\{ \a \b \g \}q} +
{2 \over 3} \e_{\g \{ \a} \kappa^i_{\b \}q} ,
\end{equation}
where $\{...\}$ and $[...]$ denote, respectively, the symmetrization and
antisymmetrization of the indices enclosed.

Comparing \p{5.6} with \p{5.7} in \p{5.4} we have
\begin{equation}\label{5.8}
 \d_{qp} \psi_{\{ \a \b \g \}\dot r}
=
2\kappa^{i}_{\{ \a \b \g \}\{ q}
\g^i_{p\} \dot r} ,
\end{equation}
\begin{equation}\label{5.9}
 \d_{qp} \e_{\g \{ \a }
\psi_{ \b \}\dot r} = - 2\e_{\g \{ \a}
\kappa^{i}_{\b \}\{ q} \g^i_{p \} \dot r} ,
\end{equation}
\begin{equation}\label{5.10}
0 = \e_{\a \b} \kappa^{i}_{\g [q} \g^i_{p] \dot r}.
\end{equation}
Substituting the solution to \p{5.9}
\begin{equation}\label{5.11}
\kappa^{~~i}_{\a q} =
- 2\g^i_{q \dot q}
\psi_{\a \dot q}
\end{equation}
into \p{5.10} we get
\begin{equation}\label{5.12}
\psi_{\a \dot q}
\tilde{\g}^i_{\dot q [q}
\g^i_{p] \dot r} = 0,
\end{equation}
which has only the trivial solution
\begin{equation}\label{5.13}
\psi_{\a \dot q}
\equiv
2i\g^{a~\b}_{~\a} D_a \Theta^{\underline{\mu}}
v_{\underline{\mu}~\b \dot q}  = 0.
\end{equation}
Eq. \p{5.13} is just the equation of motion of the Grassmann superfield
$\Theta^{\underline{\mu}}$. Due to \p{5.11} we also have
\begin{equation}\label{5.14}
\kappa^{~~i}_{\a q} \equiv
{1\over 4}\g^{a~\b}_{~\a}\Omega^{~~ai}_{\b q}
=  0 ,
\end{equation}
and, as a consequence of the supersymmetric Codazzi equation \p{4.25}
(or \p{k12}) and
the condition \p{5.30}
\begin{equation}\label{5.15}
\Omega^{~ai}_{a} \equiv
E^{~M}_{a} \Omega^{~ai}_{M} = 0 ,
\end{equation}
which is, in fact, the equation of motion for $X^{\underline{m}}$
superfield.

{}From eqs. \p{5.7}, \p{5.14} it follows that
\begin{equation}\label{5.17}
{1\over 4}\Omega^{~ai}_{\a q} =
i \g^{a~\b \g} \psi_{\{ \a \b \g\}\dot q}
\tilde{\g}^i_{\dot{q} q}.
\end{equation}
On the contrary, if $
\Omega^{~ai}_{\a q}$ has the form \p{5.17},
eq. \p{5.14} is identically satisfied due to the properties of the
 $\g$--matrices
in $d=3$ (see Appendix {\bf A}).

Thus in the case of the N=1, D=11 supermembrane
the geometrodynamical condition \p{4.6} determines the minimal embedding
of the world supersurface into the flat superspace.

In other words, the geometrodynamical condition
\p{4.6} leads to the equations of motion of the twistor-like
N=1, D=11 supermembrane, and, hence, as has been pointed out by Galperin
and Sokatchev \cite{gs2}, when one introduces the geometrodynamical
condition into a twistor--like supermembrane action with a Lagrange
multiplier one may encounter the problem with eliminating redundant propagating
degrees of freedom of the Lagrange multiplier
\cite{gs2,tp93,bers94}\footnote{This point has been missed in \cite{tp93}}.

We conclude that in the framework of the geometrical approach
the dynamics of the
N=1, D=11 supermembrane is described by internal geometry on the world
supersurface (i.e. d=2+1, n=8 supergravity) subject to the
constraints \p{4.1}, and by the superfields
 $\psi_{\{\a\b\g\}\dot q}$ \p{5.17} satisfying the supersymmetric
counterparts \p{4.16}--\p{4.18} of the Codazzi, Gauss and Ricci
equation, which determine the minimal embedding of the world
supersurface into the target superspace.
 With respect to the supergravity on the world
supersurface $\psi_{\{\a\b\g\}\dot q}$ can be regarded as the matter
superfields.

\chapter{$D=10$ superstrings}
\section{Type II superstrings}
In the case of twistor--like type II superstrings in D=10 the situation
is the same as in the N=1, D=11 supermembrane, i.e. the geometrodynamical
condition \p{4.6} causes the strings to be on the mass shell and the
embedding of a d=1+1, n=(8,8) worldsheet superspace into target
superspace is minimal. The proof is almost straightforward for a
type IIA superstring, since it can be obtained from the N=1, D=11
supermembrane by the dimensional reduction. For a D=10 IIB
superstring, characterized by Grassmann Majorana--Weyl coordinates
$\Th^{1\underline\mu}$, $\Th^{2\underline\mu}$ of the same chirality,
solving for and getting the consequences of the twistor constraint
\p{4.9} can be performed along the lines of ref. \cite{gs2} for an N=2, D=3
twistor--like superstring, but using the Lorentz harmonics allows one to
do this in a Lorentz covariant way.

Below we consider the consequences of the geometrodynamical condition
and some features of the D=10 twistor--like IIA,B superstrings in the
geometrical approach.

\subsection{Lorentz harmonics in D=10}
The spinor Lorentz harmonics which determine a local frame in a flat
 D=10 target superspace have the following form
\begin{equation}\label{6.6}
v^{~\underline{\a}}_{\underline{\mu}} =
(v^{~~+}_{\underline{\mu}~q}~,~v^{~~-}_{\underline{\mu}~\dot q} )
\qquad \in  \qquad Spin(1,9)
\end{equation}
and the inverse harmonics are:
\begin{equation}\label{6.7}
v_{\underline{\a}}^{~\underline{\mu}} =
\left(
\matrix{
v^{-~\underline{\mu}}_{q} \cr
v^{+~\underline{\mu}}_{\dot q} \cr}
\right)
\qquad \in  \qquad Spin(1,9)
\end{equation}
\begin{equation}\label{6.8}
v^{~\underline{\b}}_{\underline{\mu}}
v^{~\underline{\nu}}_{\underline{\b}} =
\d^{~\underline\nu}_{\underline\mu} , \qquad
v^{~\underline{\mu}}_{\underline{\a}}
v^{~\underline{\b}}_{\underline{\mu}} =
\left(\matrix{
\d_{qp} & 0 \cr 0 & \d_{\dot{q} \dot{p}} \cr} \right)
\end{equation}
($(+,-)$ stand for the  spinor indices,
 while their pairs $(--,++)$ stand for the vector indices of SO(1,1) in a
light--cone basis, and ${\underline\mu},{\underline\nu}=1,...,16$).

Because of the absence of the matrix of the charge conjugation of the
Majorana--Weyl spinors in D=10 there is no direct linear expression of
\p{6.7} in terms of \p{6.6} and vice versa.

The local vector frame
$
u^{\underline{a}}_{\underline{m}} =
(u^{--}_{\underline{m}},~
u^{++}_{\underline{m}},~
u^{~i}_{\underline{m}})
$
can be composed either of \p{6.6}
\begin{equation}\label{6.9}
\d_{qp} u^{++}_{\underline{m}} =
v^{+}_{q} \tilde{\G}_{\underline{m}} v^{+}_{p} ,
\end{equation}
\begin{equation}\label{6.10}
\d_{\dot{q} \dot{p}} u^{--}_{\underline{m}} =
v^{-}_{\dot q} \tilde{\G}_{\underline{m}} v^{-}_{\dot p} ,
\end{equation}
\begin{equation}\label{6.11}
\g^{i}_{q \dot{p}} u^{i}_{\underline{m}} =
v^{+}_{q} \tilde{\G}_{\underline{m}} v^{-}_{\dot p} ,
\end{equation}
or \p{6.7}
\begin{equation}\label{6.12}
\d_{qp} u^{--}_{\underline{m}} =
v^{-}_{q} \G_{\underline{m}} v^{-}_{p} ,
\end{equation}
\begin{equation}\label{6.13}
\d_{\dot{q} \dot{p}} u^{++}_{\underline{m}} =
v^{+}_{\dot q} \G_{\underline{m}} v^{+}_{\dot p} ,
\end{equation}
\begin{equation}\label{6.14}
- \g^{i}_{q \dot{p}} u^{i}_{\underline{m}} =
v^{-}_{q} \G_{\underline{m}} v^{+}_{\dot p}.
\end{equation}

%\begin{eqnarray}\label{6.16}
%u^{--}_{\underline{m}}
%u^{--\underline{m}} = 0 =
%u^{++}_{\underline{m}},~
%u^{++\underline{m}} , \qquad
%\nn
%u^{\pm \pm}_{\underline{m}}
%u^{i\underline{m}} = 0 , \qquad \nn
%u^{--}_{\underline{m}}
%u^{++\underline{m}} = 2 \qquad \nn
%u^{~i}_{\underline{m}}
%u^{j\underline{m}} = - \d^{ij} . \qquad
%\end{eqnarray}

The Lorentz harmonics $v^{~\underline{\a}}_{\underline{\mu}}$ or
$u^{\underline{a}}_{\underline{m}}$ parametrize a coset space
${{SO(1,9)}\over{SO(1,1)\times SO(8)}}$. Note that if only half of the
harmonics (for example, $v^{-~\underline{\mu}}_{q}$)
is involved in the description of a model, boost transformations of the
form
\begin{equation}\label{6.17}
\d v^{-~{\underline \mu}}_{q} = 0 ,  \qquad
\d v^{+~{\underline \mu}}_{\dot q}
= b^{++i} \g^i_{q \dot q}
v^{-~{\underline \mu}}_{q},
\end{equation}
\begin{equation}\label{6.18}
\d u^{--}_{\underline{m}} = 0 , \qquad
\d u^{++}_{\underline{m}}
= b^{++i} u^{i}_{\underline{m}} , \qquad
\d u^{i}_{\underline{m}}
= {1 \over 2} b^{++i} u^{--}_{\underline{m}} , \qquad
\end{equation}
become a symmetry of the model and can be used for reducing a number
of independent variables in \p{6.7}, \p{6.12}--\p{6.14}
to that which parametrize an $S^8$ sphere
being a compact subspace of ${{SO(1,9)}\over{SO(1,1)\times SO(8)}}$.
This is the case of a twistor--like formulation of an N=1 heterotic
string \cite{dghs92}, while a complete twistorization of the model
\cite{bcsv}
restores the coset space ${{SO(1,9)}\over{SO(1,1)\times SO(8)}}$.

\subsection{Geometrodynamical condition and twistor constraint for
type II superstrings}

One may consider n=(8,8) worldsheet superspace, where odd supervielbein
components belong either to the
same ($E^{+q},~E^{-q}$) or the different ($E^{+q},~E^{-\dot
q}$) spinor representations of $SO(8)$. The former case is appropriate for
the IIA superstring obtained by the dimensional reduction of the N=1, D=11
supermembrane \cite{tp93}, while for a IIB superstring we choose the latter
case. In both cases we assume the worldsheet supergravity constraints
\p{4.1} to imply \begin{equation}\label{ws}
\{D_{-p},D_{-q}\}=2i\d_{pq}D_{--},
\qquad
\{D_{+p},D_{+q}\}=2i\d_{pq}D_{++},
\qquad
\{D_{-p},D_{+q}\}=R_{-p+q}.
\end{equation}

The embedding in question of the $n=(8,8)$
worldsheet superspace into the flat $D=10$
\begin{equation}\label{6.26}
IIA~~~ Z^{M} = (X^{\underline m}, \Theta^{\underline{\mu}1} ,
\Theta^{2}_{\underline \mu}), \qquad or \qquad
IIB~~~Z^{M} = (X^{\underline m}, \Theta^{\underline{\mu}1} ,
\Theta^{\underline{\mu}2})
\end{equation}
target superspace is specified by the geometrodynamical condition
\begin{equation}\label{6.27}
IIA: \qquad
\Pi^{~\underline m}_{\pm q} =
D_{\pm q} X^{\underline m}
- i D_{\pm q} \Theta^{1} \G^{\underline m} \Theta^{1}
- i D_{\pm q} \Theta^{2} \tilde{\G}^{\underline m} \Theta^{2}=0
, \qquad
\end{equation}
or
\begin{equation}\label{6.28}
IIB: \qquad
\cases {
{\Pi^{~\underline m}_{+q} =
D_{+q} X^{\underline m}
- i D_{+q} \Theta^{1} \G^{\underline m} \Theta^{1}
- i D_{+q} \Theta^{2} \G^{\underline m} \Theta^{2}
= 0} \cr
{\Pi^{~\underline m}_{-\dot q} =
D_{-\dot q} X^{\underline m}
- i D_{-\dot q} \Theta^{1} \G^{\underline m} \Theta^{1}
- i D_{-\dot q} \Theta^{2} \G^{\underline m} \Theta^{2}
= 0} \cr}.
\end{equation}

The twistor constraints, which follow from \p{6.27}, \p{6.28} and \p{ws},
are
\begin{equation}\label{6.31}
IIA: \qquad
\d_{qp} (\g^{++}_{\a \b} \Pi^{~\underline m}_{++}+
\g^{--}_{\a \b} \Pi^{~\underline m}_{--}) =
D_{\a q} \Theta^{1} \G^{\underline m} D_{\b p} \Theta^{1} +
D_{\a q} \Theta^{2} \tilde{\G}^{\underline m} D_{\b p} \Theta^{2}
, \qquad
\end{equation}
\begin{equation}\label{6.32}
IIB: \qquad
\cases {
{\d_{qp} \Pi^{~\underline m}_{++} =
D_{+q} \Theta^{1} \G^{\underline m} D_{+p} \Theta^{1}
+D_{+q} \Theta^{2} \G^{\underline m} D_{+p} \Theta^{2} ,
} \cr
{\d_{\dot{q} \dot{p}} \Pi^{~\underline m}_{--} =
D_{-\dot q} \Theta^{1} \G^{\underline m} D_{- \dot p} \Theta^{1}
+ D_{-\dot q} \Theta^{2} \G^{\underline m} D_{- \dot p} \Theta^{2}
} \cr
{D_{-\dot q} \Theta^{1} \G^{\underline m} D_{+p} \Theta^{1}
+ D_{-\dot q} \Theta^{2} \G^{\underline m} D_{+ p} \Theta^{2}
= 0} \cr}
\end{equation}
$
(\g^{++}_{\a \b} \equiv \d^{+}_{\a} \d^{+}_{\b} ,
{}~~\g^{--}_{\a \b} \equiv \d^{-}_{\a} \d^{-}_{\b})
$.

By performing the dimensional reduction of the supermembrane relation
\p{5.31}, or by direct computation (Appendix B)
one gets the general solution to the
type IIA superstring twistor constraints \p{6.31} in the form
\begin{eqnarray}\label{6.35}
IIA: \qquad
&D_{+q} \Theta^{\underline{\mu} 1} =
v^{\underline{\mu}}_{+q} , \qquad
D_{+q} \Theta_{\underline{\mu} 2} = 0 , \qquad
\nn
&D_{-q} \Theta^{\underline{\mu} 1} = 0, \qquad
D_{-q} \Theta^{2}_{\underline{\mu}} =
v_{-q\underline{\mu}} , \qquad
\end{eqnarray}
and to the type IIB twistor constraints \p{6.32} (see
\cite{gs2} for the N=2, D=3 superstring) in the form
\begin{eqnarray}\label{6.36}
IIB: \qquad
&D_{+q} \Theta^{\underline{\mu} 2} =
- \tan \phi D_{+q} \Theta^{\underline{\mu} 1} =
- \sin\phi
v^{\underline{\mu}}_{+q} , \qquad
\nn
&D_{- \dot q} \Theta^{\underline{\mu} 1} =
\tan \phi  D_{- \dot q} \Theta^{\underline{\mu} 2} =
\sin\phi
v^{\underline{\mu}}_{-\dot q}.
\end{eqnarray}
For both cases
\begin{equation}\label{6.37}
\Pi^{~{\underline m}}_{++}
= u^{\underline m}_{--}, \qquad
\Pi^{~{\underline m}}_{--}
= u^{\underline m}_{--} , \qquad
\end{equation}
and the Virasoro conditions
\begin{equation}\label{6.38}
(\Pi^{~{\underline m}}_{++})^{2} = 0 =
(\Pi^{~{\underline m}}_{--})^{2}
\end{equation}
are identically satisfied. In \p{6.35}--\p{6.37} it is implied that the
scale factor $W$ is gauged away by the super--Weyl symmetry \p{weyl},
and
$\phi$ is a superfield parameter of the $SO(2)$ rotations
which mix
$\Theta^{\underline{\mu} 1}$ and
$\Theta^{\underline{\mu} 2}$.
The presence of this parameter distinguishes the $IIB$ case from
the $IIA$ one where such mixing is impossible because of the
different chirality of $\Theta^{1}$ and $\Theta^{2}$. Further analysis
shows that $\phi$ is to be a constant.

Indeed, using \p{ws}
we derive the selfconsistency conditions  for \p{6.36}
\begin{equation}\label{6.42}
-2i \d_{qp} (D_{++} \Theta^{\underline{\mu} 2} +
\tan \phi D_{++} \Theta^{\underline{\mu} 1}) =
- D_{+\{ p} \tan \phi D_{+ q \} } \Theta^{\underline{\mu} 1}
\end{equation}
\begin{equation}\label{6.43}
-2i \d_{\dot{q}\dot{p}} (D_{--} \Theta^{\underline{\mu} 1} +
\tan \phi D_{--} \Theta^{\underline{\mu} 2}) =
- D_{-\{ \dot{p} } \tan \phi D_{\dot{q} \} }
\Theta^{\underline{\mu} 2}.
\end{equation}
Contracting \p{6.42} with  $v_{-r\underline{\mu}}$ and
taking into account Eq. \p{6.36} we get
$$
\cos\phi\d_{r \{ p} D_{+q \} } \tan \phi =
-2i \d_{\dot{q}\dot{p}} (D_{--} \Theta^{\underline{\mu} 1} +
\tan \phi D_{--} \Theta^{\underline{\mu} 2})v_{-r\underline{\mu}}.
$$
{}From which it follows that for any $r=p \not= q$
\begin{equation}\label{6.44}
D_{+q } \tan \phi = 0 , \qquad \rightarrow \qquad
D_{++} \tan \phi = 0
\end{equation}
Following the same reasoning from \p{6.43} we get
\begin{equation}\label{6.45}
D_{- \dot q } \tan \phi = 0 , \qquad \rightarrow \qquad
D_{--} \tan \phi = 0.
\end{equation}
 Thus,
\begin{equation}\label{6.46}
\phi =~const~,
\end{equation}
and one may choose  new Grassmann coordinates $\hat\Th^{1,2}$ as a linear
combination of the old ones in such a way that the new variables satisfy the
chirality conditions analogous to \p{6.35}
\begin{eqnarray}\label{6.47}
IIB: \qquad
D_{- \dot q}\hat\Theta^{\underline{\mu} 1}\equiv
D_{- \dot q}(\cos\phi\Theta^{\underline{\mu} 1} -
\sin\phi  \Theta^{\underline{\mu} 2}) = 0 ,
\qquad\nn
D_{+q}\hat\Th^{\underline{\mu} 2}\equiv D_{+q} (\cos\Theta^{\underline{\mu} 2}
+
\sin\phi  \Theta^{\underline{\mu} 1}) =
0. \qquad
\end{eqnarray}
Note that the SO(2) rotations of $\Th^{1,2}$ are not a symmetry of the
IIB superstring \cite{gschw84,gsw}.

 From  \p{6.35}, \p{6.47} it follows that
\begin{equation}\label{6.48}
D_{--} \Theta^{\underline{\mu}1} = 0=D_{--}\hat\Theta^{\underline{\mu} 1},
\qquad
D_{++} \Theta^{2}_{\underline{\mu}} = 0=D_{++}\hat\Th^{\underline{\mu} 2} ,
\end{equation}
which are evidently dynamical equations.
Hence, the geometrodynamical condition \p{6.31} leads to equations of motion
of the $D=10$ type II superstrings (see \cite{gs2}) and taking into account
the results of Section {\bf 2.3}, we  conclude that the
embedding of the worldsheet n=(8,8) superspaces into the flat
$D=10,~N=2$  superspaces is minimal.

In the next subsection we will present the set of variables describing the
dynamics of the $D=10$ IIA superstring in the geometrical approach.

\subsection{Geometrical description of the $D=10$, IIA superstring}

As we have already noted the most direct and simplest way to analyze
the particular
features of the D=10 IIA superstring in the geometrical approach is to
perform dimensional reduction of the D=11 supermembrane equations
from Chapter 3, and we only declare the results.

\begin{description}
\item[i)]
superfield $W$ can be eliminated either with the use of the super--Weyl
symmetry, or as a result of equations of motion;
\item[ii)]
the induced $SO(1,1) \times SO(8)$ connection completely
coincides with the intrinsic one (compare with \p{0}), i.e.
$
\Omega^{++--}(D) = 0 ,~~
\Omega^{ij}(D)= 0,
$
so that the internal properties of the worldsheet superspace are described
by d=1+1, n=(8,8) supergravity subject to the constraints \p{ws};
\item[iii)]
the superstring modes transversal to the worldsheet are described by the
pullback of the
${{SO(1,9)} \over {SO(1,1) \times SO(8)}}$ Cartan form
\begin{eqnarray}\label{7.7}
\Omega^{++i} &= E^{-q}2i\gamma^i_{q \dot q}
\Psi^{+++}_{\dot q} + E^{--} \Omega^{~++~i}_{--} , \nn
\Omega^{--i} &= E^{+q} { 2i} \gamma^i_{q \dot q}
\Psi^{---}_{\dot q}
+ E^{++} \Omega^{~--~i}_{++} ,
\end{eqnarray}
where $\Psi^{+++}_{\dot q},~\Psi^{---}_{\dot q}$ are (anti)chiral
Grassmann superfields
\begin{eqnarray}\label{7.3}
D_{+p} \Psi^{+++}_{\dot q} &= 0, \qquad    \Rightarrow
D_{++} \Psi^{+++}_{\dot q} = 0, \qquad \nn
D_{-p} \Psi^{---}_{\dot q} &= 0, \qquad
\Rightarrow
D_{++} \Psi^{---}_{\dot q} = 0, \qquad
\end{eqnarray}
subject to  the  Codazzi--like conditions \p{4.25} (compare with \cite{howe94})
\begin{eqnarray}\label{7.4}
D_{-p} \Psi^{+++}_{\dot q} &=
{1 \over 2} \g^i_{p \dot q} \Omega^{~++~i}_{--}
, \qquad   \nn
D_{+p} \Psi^{---}_{\dot q} &=
{1\over 2} \g^i_{p \dot q} \Omega^{~--~i}_{++}
, \qquad
\end{eqnarray}
\item[iv)]
the intrinsic $SO(1,1)\times SO(8)$ curvature tensor is expressed
in terms of \p{7.7} through
the doubly supersymmetric counterparts of the
Gauss and Ricci equations \p{4.17}, \p{4.18}, \p{4.26}, \p{4.27}, for
instance,
\begin{eqnarray}\label{7.5}
R^{++--}_{-p+q} &= {2} \Psi^{+++}_{\dot p}
\Psi^{---}_{\dot q}
\g^{i}_{q \dot q}
\g^{i}_{p \dot p},  \nn
R^{ij}_{-p+q} &= {2} \Psi^{+++}_{\dot p}
\Psi^{---}_{\dot q}
\g^{[i}_{q \dot q}
\g^{j]}_{p \dot p},
\end{eqnarray}
and we see that the spinor--spinor components of the curvature tensor are
nilpotent, while the vector components are valuable (this situation
should be understood yet);
\item[v)]
upon getting the information about the geometrical objects mentioned in
items i)--iv) one may restore (with taking into account appropriate
boundary conditions) the coordinate functions
$Z^{\underline{M}}(z^M)$ from the supercovariant forms
\begin{eqnarray}\label{7.8}
 d \Theta^{\underline{\mu} 1}
&= E^{+q} v^{-~\underline{\mu} }_{q} +
E^{++} \Psi^{---}_{\dot q} v^{+~\underline{\mu} }_{\dot q}
, \qquad \nn
 d \Theta^{2}_{\underline{\mu}}
&= E^{-q} v^{~+}_{\underline{\mu} q} +
E^{--} \Psi^{+++}_{\dot q} v^{~-}_{\underline{\mu} \dot q}
, \qquad \nn
&\Pi^{\underline{m}} = E^{++} u_{++}^{\underline{m}}+
E^{--} u_{--}^{\underline{m}}. \qquad
\end{eqnarray}
\end{description}

\section{D=10 twistor--like heterotic string}
For comparison, let us make some comments on the N=1 supersymmetric part of
the twistor--like heterotic string \cite{hsstr}--\cite{bcsv} in D=10.

It is well known
that the geometrodynamical condition and the twistor constraint do not
lead to the equations of motion of the heterotic string
\cite{hsstr}--\cite{bcsv}, so
the embedding of heterotic (8,0) supersurface
$
( \xi^{\pm \pm} , \eta^{+q} )
$
into D=10, N=1 target
superspace $( X^{\underline{m}}, \Theta^{\underline{\mu}} )$
specified solely by the geometrodynamical condition is
non--minimal.

The geometrodynamical condition and the twistor constraint are obtained
from \p{6.27}--\p{6.32} by keeping only (+) SO(1,1) indices and have the
following form, respectively:
\begin{equation}\label{6.1}
\Pi^{~\underline{m}}_{+q} \equiv
D_{+q} X^{\underline{m}}
- i D_{+q} \Theta \G^{\underline{m}} \Theta = 0 ,
\end{equation}
\begin{equation}\label{6.3}
\d_{qp} \Pi^{~\underline{m}}_{++} =
D_{+q}\Theta \G^{\underline{m}} D_{+p} \Theta .
\end{equation}
A consequence of \p{6.3} is one of the Virasoro conditions \p{6.38}
\begin{equation}\label{6.4}
\Pi^{~\underline{m}}_{++} \Pi_{++~\underline{m}} = 0.
\end{equation}
In the twistor-like formulation of refs. \cite{hsstr}--\cite{bstv}, as in the
conventional Green--Schwarz formulation,
the second Virasoro condition follows from varying an action with
respect to the vielbeins, and hence {\it a priori} is not related to
another twistor constraint.

If we work within this version then only half of the Lorentz spinor
harmonics \p{6.6}, \p{6.7} are involved, since upon performing
an appropriate gauge fixing (see previous sections and Appendix B)
one gets from \p{6.3} that
\begin{equation}\label{6.19}
D_{+q} \Theta^{\underline{\mu}} =
v^{\underline{\mu}}_{+q} ,
\end{equation}
\begin{equation}\label{6.20}
\Pi^{~{\underline m}}_{++} = u_{++}^{{\underline
m}}={1\over 8}v_{+q}\G^{\underline m}v_{+q},
\end{equation}
but now we do not have any restrictions on $D_{--}\Th^{\underline\mu}$,
and
$v^{\underline{\mu}}_{-\dot q}$ never appears in this version
\cite{dghs92}.
Hence, as we have noted in Subsection 4.1.1, such a model is invariant
under the eight--parameter boost
symmetry \p{6.17}, \p{6.18} which allows one to reduce a number of
independent variables contained in $v^{\underline{\mu}}_{+q}$ to that
parametrizing an $S^8$ sphere \cite{dghs92}. This
symmetry is broken by requiring
\begin{equation}\label{6.21}
\Pi^{~{\underline m}}_{--} u^{~i}_{\underline m} = 0 ,
 \end{equation}
which, in assumption that $\Pi_{\underline{m}
--}\Pi_{--}^{\underline m}=0$ and $\Pi_{--}\Pi_{++}\not =0$,
 implies (see eqs. \p{6.12}--\p{6.14})
\begin{equation}\label{com}
\d_{\dot p\dot q}\Pi^{~{\underline m}}_{--}=\d_{\dot q\dot p}
u_{--}^{\underline{m}} =
v_{-\dot q}\tilde{\G}^{\underline{m}} v_{-\dot p}.
\end{equation}
As a result all the spinor Lorentz
harmonics become involved in to the game, and the relevant coset space
is to be ${{SO(1,9)}\over{SO(1,1)\times SO(8)}}$.

If one requires eq. \p{com} to be obtained from a heterotic string
action functional, than one gets a completely twistorized heterotic
string formulation considered in \cite{bcsv}. And it is just this version
which is more appropriate for developing the geometrical approach in the
framework discussed herein.

Thus, the embedding of the heterotic worldsheet into the flat target
superspace is described by the following pullback of the supercovariant
forms $d\Th^{\underline\mu},~\Pi^{\underline m}$
\begin{equation}\label{6.22}
d \Theta^{\underline \mu} =
E^{+q} v^{-{\underline \mu}}_{q}  +
E^{++} D_{++}\Th^{\underline\mu}+
E^{--}D_{--}\Th^{\underline\mu},
\end{equation}
\begin{equation}\label{6.23}
\Pi^{\underline m} =
E^{++} u_{++}^{\underline m} +
E^{--}u_{--}^{\underline m}
\end{equation}
and by the Maurer--Cartan equations \p{4.16}--\p{4.18} for
$\Om^{\underline{ab}}(D)$ constructed from the Lorentz harmonics.
As in the case of
the supermembrane and the type II superstrings, the selfconsistency
conditions for \p{6.22}, \p{6.23}
\begin{equation}\label{6.24}
d \Pi^{\underline m}= -i d\Theta \G^{\underline m} d\Theta ,
\end{equation}
\begin{equation}\label{6.25}
dd\Theta = 0 ,
\end{equation}
may further restrict the form of $\Om^{\underline{ab}}(D)$, but one may
convince oneself that without requiring for $\Th^{\underline{\mu}}$ and
$X^{\underline{m}}$ to satisfy additional equations obtained from the
twistor--like heterotic string action \cite{dghs92} or \cite{bcsv},
the embedding is non-minimal ($\Om^{ai}_a\not=0$). The detailed
consideration of the heterotic case is beyond the scope of  the present
article.

{\large \bf{Conclusion}}

\medskip
We have performed a generalization of the geometrical approach to
describing extended objects for studying the doubly supersymmetric
twistor--like formulation of super--p--branes. Some basic features of
embedding world supersurface into target superspace specified by the
geometrodynamical condition \p{4.6} have been considered. It has been shown
that the main attributes of the geometrical approach, such as the second
fundamental form and extrinsic torsion of the embedded surface, and
the Codazzi, Gauss and Ricci equations,  have their doubly
supersymmetric counterparts. At the same time the embedding of
supersurface into target superspace has its particular features. For
instance, in general, intrinsic and induced geometry on the supersurface
may not directly coincide (for a chosen set of intrinsic geometry
constraints), though they are related to each other by means of the
geometrical equations, and the embedding may cause more rigid
restrictions on the geometrical properties of the supersurface. This has
been demonstrated with the examples of the N=1 twistor--like
supermembrane in D=11 and the type II superstrings in D=10,
where the geometrodynamical condition caused the embedded
supersurface to be minimal and puts the theories on the mass shell. This
feature seems to be related to the general problem of constructing
off--shell superfield actions for  models with the number of
supersymmetries exceeding some ``critical'' value. In the cases
considered world supersurface possesses n=(8,8) local supersymmetry
which is indirectly related to an N=4 supergravity model in D=4 by
dimensional reduction. And it is known that D=4, N=4 supergravity
constraints put the theory on the mass shell. In the case of the
twistor--like heterotic string (Section 4.2), where there are twice less
supersymmetries on the world supersurface, the off--shell superfield
formulations do exist \cite{hsstr}--\cite{bcsv}.

Preliminary studies of N=2 twistor--like superparticles and N=2
superstrings in D=4 (with n=(2,2) worldsheet supersymmetry) in a version
close to that of refs. \cite{dis} also show that the geometrodynamical
condition does not result in equations of motion, and one may hope to
write down a superfield action without facing the problem of propagating
undesirable degrees of freedom.

As it was mentioned in the Introduction in the present paper we have not
discussed the role of the Wess--Zumino term
\cite{gschw84,gsw,hsstr,dghs92}.
The place of the Wess-Zumino form
in the geometrical approach is to be understood yet, and we shall only
make one comment. As we have seen, in the geometrical approach a basic
role is played by the Maurer--Cartan equations for the one--forms
determining the Lorentz group $SO(1,D-1)$ (eqs. \p{2.42}--\p{2.44}) and
for the supercovariant one--forms on the target superspace (eqs. \p{3.1},
\p{4.7}, \p{4.8}). But in multidimensional curved target superspace
supergravity is also characterized by a Grassmann antisymmetric, so called
Kolb--Ramond, superfield. And it is just this superfield and its
curl that contribute to the components of the Wess--Zumino form. The
Kolb--Ramond superfield acquires geometrical meaning in a generalized
group--manifold approach originated from a D=11 supergravity paper
\cite{free}
\footnote{see also \cite{town} for the case relevant to superstrings and
supermembranes}, where generalized Maurer--Cartan equations for
higher--degree differential forms (such as the Kolb--Ramond superfield) were
proposed.  Taking into consideration these generalized Maurer--Cartan
equations together with eqs. \p{4.7}, \p{4.8} should involve the
Wess-Zumino form into the geometrical approach.

Beside the main purpose of the paper concerning the geometrical
approach, we have also tried to demonstrate that the twistor--like
spinors and the spinor Lorentz harmonics are closely related to each
other  and both describe the components of the local frame in
target superspace.
In this respect the Lorentz--harmonic formulation of super--p--branes
developed in \cite{bh}--\cite{bz0} can be regarded as a component version of
the
superfield twistor--like approach. The former is based on an action
analogous to eq. \p{2.1} (with $u^a_{\underline m}$ being composed of
harmonic (or twistor) components) and seems to be related to the
geometrical approach in the most direct way. Thus if one tries to find
some dynamical ground for developing the doubly supersymmetric
geometrical approach to super--p--branes, it seems reasonable to
construct a superfield generalization of the action \p{2.1} or \p{2.1a}.
An example of such an action for N=1 massless superparticles in D=3,4 and
6 has been considered in \cite{bnsv}.

\bigskip
\bigskip
{\bf Acknowledgements}

P.P, D.S. and M.T. are grateful to N. Berkovits, P. Howe and E.
Sokatchev for fruitful discussion at various stages of this work.
%%%%%%%%%%%%%%%%%%%%
%\setcounter{appendix}A
\appendix{\bf Appendix A: NOTATION and CONVENTION}\label{A}

\begin{center}
{\bf $D=11$ $\Gamma$--matrices,
and spinor moving frame attached to\\
 supermembrane world supersurface}
\end{center}

For describing the $D=11$ supermembrane we use  the  following
$SO(1,2)\times SO(8)$
invariant splitting of the charge conjugation matrix and
the $\Gamma $-matrices in $D=11$.
\begin{eqnarray}\label{A.1}
C^{\underline{\a}\underline{\b}}= - C^{\underline{\b}\underline{\a}}= &
{\it diag} \left(\epsilon ^{\a \b}\delta_{qp},~
- \epsilon_{\a \b} \delta_{\dot{q}\dot{p}}\right),
\nn
\nn
C^{-{ 1}}_{\underline{\a}\underline{\b}} =&
\hbox{ {\it diag}}
\left( \epsilon _{\a \b} \delta _{qp} ,~
- \epsilon ^{\a\b} \delta_{\dot{q}\dot{p}} \right) ,
\nn
\nn
&(\Gamma^{\underline{a}})^{~\underline{\b}}_{\underline\a}
 \equiv \left(\Gamma^{a}, \Gamma ^{i} \right),
\nn
\nn
(\Gamma ^{a})^{~\underline{\b}}_{\underline\a} \equiv
\left(\Gamma ^{\underline{ 0}},~\Gamma^{\underline{ 9}},
{}~\Gamma^{\underline{ 10}} \right)
\equiv
\left(\Gamma ^{{ 0}},\Gamma ^{{ 1}},\Gamma ^{{ 2}}\right)
=& {\it diag} \left( \gamma ^{a~ \b}_{~\a} \delta _{qp},
- \gamma ^{a~\a}_{\b} \delta _{\dot{q} \dot{p}} \right),
\nn
\nn
(\Gamma ^{i})^{~\underline{\b}}_{\underline\a} \equiv
\left(\Gamma ^{1},\ldots , \Gamma ^{8}\right)=
&\left[ \matrix{ 0 & \epsilon _{\a \b} \gamma ^{i}_{{q}\dot{p}} \cr
- \epsilon ^{\a \b} \tilde{\gamma}^{i}_{{\dot q}{p}} \cr}
\right]
\end{eqnarray}
%%%%%%%%%%%%%%%%%%%%
where
$
\underline{\a} = \left(^{\a}_{q},~ _{\a {\dot q}} \right)
$
is composed of $SO(1,2)\times SO(8)$ spinor indices,
$\underline\mu,\underline\a =1,~\ldots ,~32$
are  the  spinor  indices of
$SO(1,10)$; $\a, \b = ~1, ~2$ are the spinor indices of
$SO(1,2)$, and $q,~p = 1,\ldots ,8$ ; ${\dot q} ,~{\dot p} =1~,~\ldots ,~8$ are
$s$- and $c$- spinor indices  of  SO(8), respectively;
$~~~\tilde{\gamma }^{i}_{{\dot q}{p}}
\equiv  \gamma ^{i}_{p\dot q}$
are $d=8$ $\gamma $-matrices,
$~~~\gamma ^{a~\a}_{~\b}$  are
$d=3$
$\gamma $--matrices,
$\epsilon ^{{{\a}{\b}}}=- \epsilon
^{{{\b}{\a}}}$
$(\epsilon ^{{ 12}}= \epsilon _{{ 12}}= i)$ is the
$d=3$ charge conjugation matrix.

Note that $D=11$ as well as $d=3$ gamma matrices
with both indices being upper or lower are
symmetric:
\begin{eqnarray}\label{A.sim}
\left(\Gamma ^{\underline{ m}}C^{-{ 1}}\right)^{{ T}} =
\left(\Gamma ^{\underline{ m}}C^{-{ 1}} \right) , \nn
\left(\epsilon \gamma ^{a}\right)^{{ T}} =
\left(\epsilon \gamma ^{a}\right).
\end{eqnarray}
In the main text and below we skip $C$ and $\e$ in the formulas like
\p{A.sim} and write, for example $\left(\Gamma ^{\underline{ m}}C^{-{ 1}}
\right)_{\underline{\a}\underline{\b}}
\equiv \left(
\Gamma ^{\underline{ m}}\right)_{\underline{\a}\underline{\b}}$.

The Lorentz harmonics form a $32 \times 32$
matrix $v^{{ {\underline\a}}}_{{\underline\mu}}$ of $Spin(1,10)$:
\begin{equation}\label{A.3}
v^{~\underline{\a}}_{\underline{\mu}}=
\left( v^{~\a}_{\underline{\mu}q},~
v_{\underline{\mu}~\a \dot q} \right).
\end{equation}

We use the ``left'' action of the charge conjugation matrices for rising
and lowering spinor indices in D=11 and d=3:
\begin{equation}\label{A.-}
v^{\underline{\mu}}_{~\b q} \equiv
C^{\underline{\mu}\underline{\nu}} v_{\underline{\nu} \b q} ,
\qquad
v_{\underline{\mu}~\b q} \equiv
C^{-1}_{\underline{\mu}\underline{\nu}} v^{\underline{\nu}}_{~ \b q} ,
\qquad
v^{\underline{\mu}~\a}_{~~q} \equiv
\e^{\a\b}
v^{\underline{\nu}}_{~ \b q} ,
\qquad
v^{\underline{\mu}}_{\a q} \equiv
\e_{\a\b}
v^{\underline{\nu} ~\b}_{~~q}.
\end{equation}

 The requirement that the matrix \p{A.3}
takes its values in the group ${\it Spin}(1,10)$ (which is the
double-covering group of $SO(1,10)$)
can be ensured by imposing the  following ``harmonic'' conditions
\begin{equation}\label{A.4}
\Xi  \equiv v^{\underline{\a}}_{\underline{\mu}}
C^{\underline{\mu}\underline{\nu}}
 v^{\underline{\b}}_{\underline{\nu}}
 - C^{\underline{\a}\underline{\b}} = 0 ,
\end{equation}
\begin{equation}\label{A.5}
\Xi ^{\underline{a}}_{\underline{ m}_{{ 1}}\underline{ m}_{{ 2}}} \equiv
v^{\underline{\a}}_{\underline{\mu}}
(\Gamma _{\underline{ m}_{{ 1}}\underline{ m}_{{ 2}}})^{\underline{\mu}
\underline{\nu}}
v^{\underline{\b}}_{\underline{\nu}}
\left(\Gamma ^{\underline{a}}
\right)_{\underline{\a}\underline{\b}} = 0 ,
\end{equation}
\begin{equation}\label{A.6}
\Xi ^{\underline{a}}_{\underline{ m}_{{ 1}}\ldots \underline{ m}_{{ 5}}} \equiv
v^{\underline{\a}}_{\underline{\mu}}
\left( \Gamma _{\underline{ m}_{{ 1}}\ldots \underline{ m}_{{ 5}}} \right)
^{\underline{\mu}\underline{\nu}}
v^{\underline{\b}}_{\underline{\nu}}
\left( \Gamma ^{{\underline{a}}}\right)_
{\underline{\a}\underline{\b}} = 0.
\end{equation}
Eqs. \p{A.4}, \p{A.-} allow one to determine the matrix
inverse to $v^{\underline{\a}}_{\underline{\mu}}$ by use of the same variables
$v^{~\a}_{\underline{\mu} q} , v_{\underline{\mu} ~\a {\dot q}}$:
\begin{equation}\label{A.7}
v^{\underline{\nu}}_{\underline{\b}} \equiv
(v^{-1})^{\underline{\nu}}_{\underline{\b}}=
\left(
- v^{\underline{\nu}}_{\b q}, v^{\underline{\nu} \b}_{~\dot q} \right).
\end{equation}

One can see that not all the relations encoded in
\p{A.4}--\p{A.6} are independent (they ``kill'' 969 = 1024-55 degrees of
freedom
\cite{bzm}). An independent subset can be chosen in different ways. For
instance, in \p{A.4} one can take the only independent condition to be
that the harmonics have unit norm. Then \p{A.5}, \p{A.6} contain
just the same information as \p{3.10}, or \p{3.15}--\p{3.17}.

Hence,  among the 1024 components of
$v^{{\a}}_{\underline{\mu} q} , v_{\underline{\mu }~\a\dot q}$
only 55=$dim~SO(1,10)$ are independent.  Among the latter
$31= 3 + 28 = {\sl dim}~SO(1,2)+{\sl dim}~SO(8)$ can be gauged away by
$SO(1,2)\times SO(8)$ local symmetry of the supermembrane theory. Thus,
$v^{{\a}}_{\underline{\mu } q} , v_{\underline{\mu }~\a\dot q}$
parametrize a coset space ${{SO(1,10)}\over{SO(1,2)\times  SO(8)}}$.

\bigskip
\begin{center}
{\bf $D=10$
$\G$--matrices, and spinor moving frame attached to\\
superstring worldsheet}
\end{center}

For $D=10$ superstrings the vector indices take ten values
$\underline{m}, \underline{a} = 0,1,\ldots ,9$
and the dimension of the Majorana--Weyl spinor representation is $16$:
$\underline{\a}, \underline{\mu} =1,\ldots ,16$.

For making computations we use the following
$SO(1,1) \times SO(8)$ invariant realization of $16\times
16$ $\G$-matrices:
\begin{equation}
\G^{\underline{ 0}}_{\underline{\a}\underline{\b}}=
\hbox{ {\it diag}}(\delta _{{ qp}},
\delta_{{\dot q}{\dot p}})
= \tilde{\G }^{\underline{0}~\underline{\a}\underline{\b}} ,
\end{equation}
\begin{equation}
\G^{\underline{9}}_{\underline{\a}\underline{\b}}=
\hbox{ {\it diag}} (\delta _{qp},
-\delta _{{\dot q}{\dot p}}) =
-\tilde{\G }^{\underline{9}~\underline{\a}\underline{\b}} ,
\end{equation}
\begin{equation}
\G ^{i}_{\underline{\a}\underline{\b}} =
\left(\matrix{0 & \gamma ^{i}_{q\dot p}\cr
\tilde{\gamma}^{i}_{{\dot q} p} & 0\cr}
\right)
= - \tilde{\G }^{i~\underline{\a}\underline{\b}} ,
\end{equation}
\begin{equation}
\G ^{{++}}_{\underline{\a}\underline{\b}} \equiv
(\G^{\underline{ 0}}+
\G ^{\underline{ 9}})_{\underline{\a}\underline{\b}}=
\hbox{ {\it diag}}(~2\delta _{qp},~ 0)
= -(\tilde{\G }^{\underline{ 0}}-
\tilde{\G }^{\underline{ 9}})^{\underline{\a}\underline{\b}} =
\tilde{\G }^{{--}~\underline{\a}\underline{\b}} ,
\end{equation}
\begin{equation}
\G ^{{--}}_{\underline{\a}\underline{\b}}\equiv
(\G ^{\underline{ 0}}-\G ^{\underline{ 9}}
)_{\underline{\a}\underline{\b}}=
\hbox{ {\it diag}}(~0, ~2\delta
_{{\dot q}{\dot p}}) = (\tilde{\G }^{\underline{ 0}}+
\tilde{\G }^{\underline{ 9}})^{\underline{\a}\underline{\b}}
= \tilde{\G }^{{++}~\underline{\a}\underline{\b}}.
\end{equation}
Note that with respect to their properties the matrices $\G,~\tilde\G$
are closer
to the D=4 Pauli matrices rather then to the Dirac matrices.

Let us also stress the absence of the charge conjugation
matrix in the
$D=10$ Majorana--Weyl spinor representation, so
the $\G$--matrices  always have both spinor
 indices down and $\tilde{\G}$ have both spinor indices up. As a result
there is no linear expression for the inverse Lorentz harmonics
($v^{\underline\mu}_{+q},~v^{\underline\mu}_{- \dot q}$)
through ($v^{+}_{\underline\mu q},~v^{~-}_{\underline{\mu} \dot q})$.
They are related by the requirement that the spinor harmonics and their
inverse define one and the same composed vector frame:
\begin{eqnarray}\label{A.9}
u^{{++}}_{\underline{m}} =&
{{1} \over {8}} v^{+}_{q} \tilde{\s}_{\underline{m}} v^{+}_{q} =
{{1} \over {8}} v^{+}_{\dot q} \s_{\underline{m}} v^{+}_{\dot q}
\qquad \nn
u^{{--}}_{\underline{m}} =&
{{1} \over {8}} v^{-}_{\dot q} \tilde{\s}_{\underline{m}} v^{-}_{\dot q} =
{{1} \over {8}} v^{-}_{q} \s_{\underline{m}} v^{-}_{q}  \qquad \nn
u^{{i}}_{\underline{m}} =&
{{1} \over {8}} \g^{i}_{q \dot q} v^{+}_{q}
\tilde{\s}_{\underline{m}} v^{-}_{\dot q} =
- {{1} \over {8}} \g^{i}_{q \dot q}
v^{-}_{q} \s_{\underline{m}} v^{+}_{\dot q}.  \qquad
\end{eqnarray}

The irreducible harmonic conditions for the matrix
($v^{+}_{\underline\mu q},~v^{~-}_{\underline{\mu} \dot q}$)
to take values in $Spin(1,9)$ are:

\begin{equation}\label{A.10}
u^{\underline{m}}_{\underline{a}}
{}~~\Xi^{\underline{a}}_{\underline{m}_{{1}}
\ldots \underline{m}_{{4}}\underline{m}}=
u^{\underline{m}}_{\underline{a}}
Sp(v^{{T}} \tilde{\G }_{\underline{m}_{{}}\ldots \underline{m}_{{4}}}
v\G ^{\underline{a}})= 0 ,
\end{equation}

\begin{equation}\label{A.11}
\Xi _{{ 0}}
\equiv  u^{--}_{\underline{m}} u^{\underline{m}{++}} - 2 = 0 ,
\end{equation}
(where eqs. \p{A.9} for $u^{\pm\pm}_{\underline m}$ are implied).
Then
$$
\Xi ^{\underline a}_{\underline{m}_{{1}} \ldots \underline{m}_{{5}} }
= 0
$$
is identically satisfied.

For the detailed discussion of the Lorentz harmonics in D=10 see refs.
\cite{zup,ghs,gds,ght,bzst,bzst1}.

%%%%%%%%%%%%%%%%%%%%%%
\appendix{{\bf Appendix B}}

Here, for the $D=11$, N=1 supermembrane and $D=10,~IIA$ superstring,
 we present a direct proof that eqs. \p{4.11}, \p{4.12} expressing
$D_{\a q}\Th^{\underline\mu}$ and $\Pi^{\underline m}_a$ in terms of the
Lorentz harmonics are the consequence of the twistor constraint \p{4.9}:
\begin{equation}
\label{twis} \d_{qp} \g^a_{~\a \b}
\Pi^{~\underline{m}}_{a} = D_{\a q}\Th
\Gamma^{\underline{m}}D_{\b p}\Th.
\end{equation}

%%%%%%%%%%%%%%%%%%%%%%%%%%%%
\noindent
 {\bf
Solving the twistor constraint for supermembrane in $D=11$.}

\bigskip
%%%%%%%%%%%%%%

Let us choose in target superspace a local moving frame
\p{3.15}--\p{3.17}:
\begin{equation}\label{B3.15}
\d_{\hat q\hat p} (\gamma_{\hat a} )_{\hat \a \hat\b}
u^{~\hat a}_{\underline{ m}} =
v_{\hat\a\hat q}
%( \Gamma_{\underline{ m}} )_{\underline{\mu } \underline{\nu}}
%v^{\underline{\nu}}_{\hat\b \hat p} ,
%\equiv v_{\hat\a\hat q}
\Gamma_{\underline{ m}}
v_{\hat\b \hat p} ,
\end{equation}
\begin{equation}\label{B3.16}
\d_{\dot{\hat q} \dot{\hat p}} (\gamma_{\hat a})^{\hat\a \hat\b}
u^{\hat a}_{\underline{ m}} =
v^{\hat\a }_{\dot{\hat q}}
\Gamma_{\underline{m}}
v^{\hat\b }_{\dot{\hat{ p}}} ,
\end{equation}
\begin{equation}\label{B3.17}
\d^{\hat\a}_{\hat\b} \gamma ^{\hat i}_{\hat q \dot{\hat{ p}}}
u^{\hat i}_{\underline{ m}}
= v_{\hat\a \hat q}
\Gamma_{\underline{ m}}
v^{\hat\b }_{\dot{\hat{ p}}}.
\end{equation}
Note that {\sl a priori} the local $SO(1,2)\times SO(8)$ group acting
from the right on the components of the local frame does not coincide
with the local $SO(1,2)\times SO(8)$  group related to the world
supersurface. This is indicated by hats on $SO(1,2)\times SO(8)$ indices.

By use of SO(1,D-1) transformations the local Lorentz frame can always
be chosen in such a way that
\begin{equation}\label{B.1}
\Pi^{\underline{m}}_{a} u^{\hat i}_{\underline{m}} = 0.
\end{equation}
Then multiplying, respectively, the l.h.s. and the r.h.s. of \p{twis}
 by the l.h.s. and the r.h.s. of \p{B3.15}--\p{B3.17} we get
\begin{equation}\label{B.4}
\d_{{qp}} \gamma^{b}_{\a \b}
{F}^{\hat{a}}_{b}=F^{\underline{\a}}_{\a q}
(\Gamma^{\hat{a}})_{\underline{\a} \underline{\b}}
F^{\underline{\b}}_{\b p},
\end{equation}
\begin{equation}\label{B.5}
F^{\underline{\a}}_{\a q}
(\Gamma^{\hat i})_{\underline{\a}\underline{\b}}
F^{\underline{\b}}_{\b p} = 0 ,
\end{equation}
where
\begin{equation}\label{B.3}
{F}^{~\hat{a}}_{{b}} \equiv
\Pi^{\underline{m}}_{b} u^{\hat a}_{\underline{m}}
\end{equation}
and
\begin{equation}\label{B.6}
F^{\underline{\a}}_{\a q} \equiv
D_{\a q} \Theta^{\underline{\mu}} v^{\underline{\a}}_{\underline{\mu}}
\equiv
(A^{~~~\hat{\a}}_{\a q~ \hat p}, B_{\a q ~\hat{\a} {\dot{\hat p}}}).
\end{equation}

Let us begin with considering eq.\p{B.5}.
Using the explicit form of the $\G$--matrices \p{A.1}
and \p{B.6} we get
\begin{equation}\label{B.7}
A^{~~\hat {\a}}_{\a q \hat p} \gamma^{\hat i}_{{\hat p}{\dot{\hat p}}}
B_{\b p \hat{\a} {\dot{\hat p}}} +
B_{\a q \hat{\a} {\dot{\hat p}}}
\tilde{\gamma}^{\hat i}_{{\dot{\hat p}}{\hat p}}
A^{~~\hat{\a}}_{\b p \hat p} = 0.
\end{equation}
Suppose that one of the  matrices, for example $A$,
is non--degenerate
\begin{equation}\label{B.8}
{\hbox {\it det}} (A^{~~~\hat{\a}}_{\a q ~\hat p}) \not= 0
\end{equation}
(otherwise the solution we would got corresponded to a null
super--p--brane (see \cite{bstv} for the string case).
Then we can rewrite \p{B.7} as follows:
\begin{equation}\label{B.9}
(A^{{-1}} B)_{\hat{\a} {\hat p} \hat{\b} {\dot{\hat q}}}
\tilde{\gamma}^{\hat i}_{\dot{\hat q}{\hat q}}
+
(A^{{-1}} B)_{\hat{\b}{\hat q}\hat{\a} {\dot{\hat q}}}
\tilde{\gamma}^{\hat i}_{\dot{\hat q}{\hat p}} = 0.
\end{equation}
Decomposing the matrix $(A^{{-1}} B)$ into the SO(1,2) irreducible parts
\begin{equation}\label{B.10}
(A^{{-1}} B)_{\hat{\a} {\hat p}\hat{\b} {\dot{\hat q}}}
\equiv
\e_{\hat{\a} \hat{\b}}  G_{{0} \hat p{\dot{\hat p}}} +
\gamma^{\hat{a}}_{\hat{\a} \hat{\b}}  G_{\hat{a} \hat p {\dot{\hat q}}} ,
\end{equation}
and substituting \p{B.10} into \p{B.9}, we get
two equations:
\begin{equation}\label{B.10.}
G_{{0} [\hat p {\dot{\hat q}}}
\tilde{\gamma}^{i}_{\dot{\hat q}{\hat q]}} = 0 ,
\end{equation}
and
\begin{equation}\label{B.11}
G_{\hat{a} \{\hat p {\dot{\hat q}}}
\tilde{\gamma}^{\hat i}_{{\dot{\hat q}}{\hat q\}}} = 0.
\end{equation}
Decomposing $G_{0}$ and $G_a$ onto
the $SO(8)$ irreducible parts
\begin{equation}\label{B.12}
G_{{0}\hat p {\dot{\hat q}}} =
G^{\hat j}_{0}  \tilde{\gamma}^{\hat j}_{\hat p {\dot{\hat{ q}}}}  ,
+ G^{\hat j\hat k \hat l}_{0}  \tilde{\gamma}^{\hat j\hat k\hat
l}_{\hat p {\dot{\hat q}}},
\end{equation}
\begin{equation}\label{B.13}
G_{{\hat{a}} \hat p {\dot{\hat q}}} =
G^{j}_{\hat{a}}  \tilde{\gamma}^{\hat j}_{\hat p {\dot{\hat q}}}  ,
+ G^{\hat j\hat k\hat l}_{\hat{a}}  \tilde{\gamma}^{\hat j\hat k\hat
l}_{\hat p {\dot{\hat q}}},
\end{equation}
and substituting \p{B.12}, \p{B.13} into  eqs. \p{B.10.}, \p{B.11} we
obtain
\begin{equation}\label{B.14}
(G^{[\hat j}_{0} \d^{{\hat k}]\hat i} + G^{[\hat j\hat k{\hat i}]}_{0})
{\tilde{\gamma}}^{\hat j\hat k}_{\hat p \hat q} = 0
,
\end{equation}
\begin{equation}\label{B.15}
G^{\hat i}_{\hat {a}} \d_{{\hat p\hat q}} +
 G^{\hat j\hat k\hat l}_{\hat{a}}
\gamma^{\hat j\hat k\hat l\hat i}_{\hat p\hat q} = 0 ,
\end{equation}
from which it follows that all the components of $G_{0}$ and $G_{\hat{a}}$
vanish, and
hence
\begin{equation}\label{B.16}
(A^{{-1}} B)_{\hat{\a}\hat p \hat{\b} {\dot{\hat q}}}
= 0 = B_{\a p \hat\b\dot{\hat q}}
\end{equation}
since the matrix $A$ was supposed to be non--degenerate.
Thus eq. \p{B.5} is valid if and only if
$$
D_{\a p}\Th^{\underline\mu}v_{\underline\mu\hat\a\dot{\hat q}}=0
$$
Let us turn to eq.\p{B.4}.
Substituting \p{B.6} and taking into account \p{B.16} we
can rewrite eq.\p{B.4} in the form
\begin{equation}\label{B.18}
A^{~~~\hat{\a}}_{\a q~\hat p}
(\gamma^{{\hat{a}}})_{\hat{\a} \hat{\b}}
A^{~~~\hat{\b}}_{\b p~\hat p}
= \d_{qp}
(\gamma^{b})_{{\a \b}} F^{\hat{a}}_{b}
,
\end{equation}
where $ F^{\hat a}_{b}$ is defined by the eqs. \p{B.1}, \p{B.3}.

To analyse eq.\p{B.18} let us expend the matrix $A$ in a complete basis of
the space of $2\times 2$ matrices:
\begin{equation}\label{B.19}
A^{~~~\hat{\b}}_{\a q~\hat p} =
a_{ q\hat p} \d^{\hat{\b}}_{\a} + b^{a}_{q\hat p}
(\gamma_{a})^{~\hat{\b}}_{\a}.
\end{equation}
Note that eq. \p{B.19} is invariant only under the diagonal subgroup
$SO(1,2)$ of the $SO(1,2)\times SO(1,2)$.

 With $A$ being in the form \p{B.19}, eq. \p{B.18} splits, in particular,
into equations
\begin{equation}\label{B.20}
(a (b^{a})^{{T}})_{{ p \hat q}} = 0,
\end{equation}
%\begin{equation}\label{B.20}
%(a (b^{a})^{{T}})_{{ \{p\hat q\} }} = 0,
%\end{equation}
%\begin{equation}\label{B.21}
%(a (b^{a})^{{T}})_{{[p\hat q] }} = 0,
%\end{equation}
which, if the matrix $a$ is nonsingular
($
{\hbox {\it det}} ~a \not= 0$), result in
\begin{equation}\label{B.23}
(b^{a})_{{p \hat q }} = 0.
\end{equation}
Then  from \p{B.18} we get
\begin{equation}\label{B.24}
(a~a^{{T}})_{{qp}} \eta^{{ a} b} = \d_{{qp}}{F}^{{a}b}.
\end{equation}
An evident consequence of \p{B.24} is that $F^{{a}b}$ is
proportional to the unit
matrix
\begin{equation}\label{B.25}
{F}^{{a}b} = W^2 \eta^{{ a}b},
\end{equation}
which results in (see \p{4.12})
\begin{equation}\label{B.26}
\Pi^{\underline{m}}_{{a}} =
W^2u_{a}^{\underline{m}},
\end{equation}
and hence
$$
\Pi^{\underline{m}}_{a}
\Pi_{\underline{m} b}=W^2 \eta_{ab}.
$$

On the other hand, eq. \p{B.24} means that the matrix $a$ is
proportional to an SO(8) matrix $L$:
\begin{equation}\label{B.27}
a_{{q\hat p}}= WL_{{q\hat p}},
\end{equation}
\begin{equation}\label{B.28}
(L L^{{T}})_{qp} = \d_{{qp}}.
\end{equation}
Now one can use the SO(8) transformations of the local frame to fix
$L_{q\hat p}=\d_{q\hat p}$, thus remaining with only one SO(8) relevant
to the supermembrane world supersurface.
As the  final result we get
\begin{equation}\label{B.31}
D_{{\a q}} \Theta^{\underline{\mu}}=
 W v^{\underline{\mu}}_{{\a} q},\qquad
\Pi^{\underline{m}}_{{a}} =
W^2u_{a}^{\underline{m}}.
\end{equation}

\noindent
{\bf
Solving the twistor constraint for type II superstrings in D=10.}

\bigskip
For the type IIA superstring this can be done either
by performing the  dimensional reduction of the
supermembrane relations \p{B.31} or by direct computation completely
analogous to that for the supermembrane case. Thus we only present the
result:
\begin{equation}\label{B2.13}
D_{+q} \Theta^{1\underline{\mu}} =
W v^{\underline{\mu}}_{+q}
, \qquad
D_{-q} \Theta^{1\underline{\mu}}= 0
; \qquad
\end{equation}
\begin{equation}\label{B2.15}
D_{+q} \Theta^{2}_{{\underline{\mu}}}= 0 ,
\qquad
D_{- q} \Theta^{2}_{{\underline{\mu}}}
= W v_{ \underline{\mu}+ q};
\end{equation}
and
\begin{equation}\label{B2.17}
\Pi^{\underline{m}}_{++}= W^2u^{\underline{m}}_{++} ,\qquad
E^{\underline{m}}_{--} = W^2 u^{\underline{m}}_{--}.
\end{equation}

The latter equations ensure the validity of the Virasoro conditions.

The (anti)chirality of $\Th^1,~\Th^2$ (eqs. \p{B2.13}, \p{B2.15}
immediately follows from \p{B.31} if one takes into account that
in D=10 $v^{\underline\mu}_{-q}$ and $v_{\underline\mu +q}$ are not
present (see Appendix A).

To convince the reader that Eqs. \p{B2.13}--\p{B2.17} correspond to the
general solution, below we indicate the main steps of the straightforward
proof. As for the supermembrane, one can choose the harmonic variables in
such a way, that
\begin{equation}\label{B.1.}
\Pi^{\underline{m}}_{a} u^{i}_{\underline{m}} = 0 , \qquad
\end{equation}
or
\begin{equation}\label{B.2.}
\Pi^{\underline{m}}_{a} = \underline{F}_{{a}{\bar{b}}}
u^{\bar{b}}_{\underline{m}} ,
\end{equation}
with
\begin{equation}\label{B.3.}
\underline{F}^{~\bar{b}}_{{a}} \equiv
\Pi^{\underline{m}}_{a} u^{b}_{\underline{m}} ,  \qquad
\end{equation}
then by use of Eqs. \p{A.9} the twistor constraints
\begin{equation}\label{B2.2}
 D_{\a q} \Theta^{1\underline{\mu}}
\G^{\underline{m}}_{\underline{\mu}\underline{\nu}}
D_{\b p} \Theta^{1\underline{\nu}}
+ D_{\a q} \Theta^{2}_{\underline{\mu}}
\tilde{\G}_{\underline{m} \underline{\mu}\underline{\nu}}
D_{\b p} \Theta^{2}_{\underline{\nu}} =
\d_{{qp}} (\gamma^{{++}}_{\a \b }
\Pi^{~\underline{m}}_{{++}}+\gamma^{{--}}_{\a \b }
\Pi^{~\underline{m}}_{{--}})
\qquad
\end{equation}
can be rewritten as follows:
\begin{equation}\label{B.4.}
F^{\underline{\hat{\a}}}_{\a q}
(\Gamma^{\bar{a}})_{\underline{\hat{\a}} \underline{\hat{\b}}}
F^{\underline{\hat{\b}}}_{\b p} = \d_{{qp}} \gamma^{b}_{\a \b}
\underline{F}^{\bar{a}}_{b} ,
\qquad
\end{equation}

\begin{equation}\label{B.5.}
F^{\underline{\hat{\a}}}_{\a q}
(\Gamma^{i})_{\underline{\hat{\a}}\underline{\hat{\b}}}
F^{\underline{\hat{\b}}}_{\b p} = 0 ,
\qquad
\end{equation}
where the  matrix $F^{\underline{\hat{\a}}}_{\a q}$ is defined by the
relation
\begin{equation}\label{B.6.}
F^{\underline{\hat{\a}}}_{\a q} =
(D_{\a q} \Theta^{1\underline{\mu}}
v^{\underline{\a}}_{\underline{\mu}},
D_{\a q} \Theta^{2}_{\underline{\mu}}
v_{\underline{\a}}^{\underline{\mu}})
=
(A^{~~~\bar{\a}}_{\a q~ A}, B_{\a q ~\bar{\a} {\dot A}})
, \qquad
\end{equation}
the indices $\hat{\underline{\a}},~\underline{\hat \b} =~1,\ldots,32$ and
$(\G^{\underline a})_{\underline{\hat \a} \underline{\hat \b}}$ are ten of
the eleven $\G$--matrices \p{A.1}. The transverse part \p{B.5.} has the
same form as eq. \p{B.5}, thus, the problem under consideration is reduced to
that having been solved for the supermembrane, and using the same
reasoning we finally arrive at the solution \p{B2.13}--\p{B2.17}.

The case of a twistor--like IIB superstring can be analyzed following
the same group--theoretical reasoning as above (see \cite{gs93} for $D=3,~
N=2$ superstring) with the result having been presented in Chapter 4.

\appendix{\bf Appendix C}

\bigskip
In which we show that for the two sets of the Maurer--Cartan equations,
namely \p{4.7}, \p{4.8} and \p{4.16}--\p{4.18}, their spinor--spinor
components ensure the validity of the rest, if the number of world
surface supersymmetries is more than one. This  means that the irreps of
the spinor--vector and vector--vector components either coincide with
that of spinor--spinor ones or can be obtained  by acting on the latter with
the spinor covariant derivatives.  Thus, for getting all the
consequences of the Maurer--Cartan equations it is sufficient to
consider just the spinor--spinor components. Of course, another choice of
the independent equations is possible.

The situation is analogous to
Bianchi identity theorems in super--Yang--Mills \cite{sohnius}
and supergravity theories \cite{Dragon}--\cite{bandzim}.

Let us consider eqs. \p{4.7},
\p{4.8} and, for convenience,  denote their l.h.s. by
$M^{\underline{M}}= (M^{\underline{m}}, M^{\underline{\mu}})$
\begin{equation}\label{C.1}
M^{\underline{m}}
\equiv {1 \over 2} E^{B} E^{C}
M^{~\underline{m}}_{CB} \equiv
d \Pi^{\underline{m}} + i d\Theta \Gamma^{\underline{m}} d\Theta=0 ,
\end{equation}
\begin{equation}\label{C.2}
M^{\underline{\mu}}
\equiv {1 \over 2} E^{B} E^{C}
M^{~\underline{\mu}}_{CB} \equiv dd\Theta^{\underline{\mu}}
= 0.
\end{equation}

Note that when $\Pi^{m}$ and $\Th^{\underline{\mu}}$ are written in
terms of the harmonics (see \p{B2.17}), equations \p{C.1}, \p{C.2}
become nontrivial.

The components of \p{C.1} and \p{C.2} in the basis of supercovariant
two--forms
$$
E^{\a q}  E^{\b p}, \qquad
E^{a} E^{\b p}, \qquad
E^{a} E^{b}  \qquad
$$
are
\begin{equation}\label{C.3}
M^{\underline{M}}_{\a q\b p}=0 , \qquad
\end{equation}
\begin{equation}\label{C.4}
M^{\underline{M}}_{\a qb}= 0 , \qquad
\end{equation}
\begin{equation}\label{C.5}
M^{
\underline{
M}}_{ab} = 0.
\end{equation}
The spinor--spinor components  (eq. \p{C.3}) are just equations
\p{4.10} and \p{4.9}.

The integrability conditions for \p{C.1}, \p{C.2}, which are analogous
to the Bianchi identities \cite{sohnius}, \cite{Dragon,GrimWZ} read
\begin{equation}\label{C.6}
I^{\underline{m}} \equiv dM^{\underline{m}} = 0 ,
\end{equation}
\begin{equation}\label{C.7}
I^{\underline \mu} \equiv dM^{\underline \mu} = 0.
\end{equation}

Suppose that eqs. \p{C.3} hold, then,
using the torsion constraints \p{4.1},
we can derive from the spinor component equations
$ I^{\underline M}_{\a q~\b p~\g r} = 0$ contained in \p{C.6},
\p{C.7}
the identity
\begin{equation}\label{C.8}
\d_{qp} \g^a_{~\a \b} M^{\underline M}_{a~\g r} +
\d_{rq} \g^a_{~\g \a} M^{\underline M}_{a~\b p} +
\d_{pr} \g^a_{~\b \g} M^{\underline M}_{a~\a q} =0
\end{equation}
For a world surface superspace with $n>1$ we can put in eq.\p{C.8}
$p = r \not= q$
for each value of $q$ and get
\begin{equation}\label{C.9}
\g^a_{~\b \g} M^{\underline M}_{\a qa} =0=
M^{\underline M}_{\a qa}.
\end{equation}
Hence, \p{C.4} is a consequence of \p{C.3}. To show that \p{C.5} is also
a consequence of \p{C.3} one should consider other components of \p{C.6},
\p{C.7} with taking into account that
$M^{\underline M}_{\a q\b p}$ and
$M^{\underline M}_{\a qb} $ vanish.

For the
$SO(1,D-1)$ Maurer--Cartan equations \p{4.16}--\p{4.18} the proof can be
performed in the same way.

In the case of n=1 world surface supersymmetry eqs. \p{C.4}, \p{C.5} may
produce independent consequences, as one may already see from eq.
\p{C.8}.


\begin{thebibliography}{99}
\bibitem{lr}
F. Lund and T. Regge \PRD 14 1976 1524.

\bibitem{om}
R.Omnes \NPB 149 1979 269.

\bibitem{barnes}
B. M. Barabashov and V. V. Nesterenko, {\sl Relativistic string model in
hadron physics}, Moscow, Energoatomizdat, 1987.

\bibitem{zhelt}
A. Zheltukhin, \SJNP 33 1981 1723 ; \TMP 52 1982 73 ;
\PLB 116 1982 147; \TMP 56 1983 230.

\bibitem{hoppe}
J. Hoppe, \PLB 250 1990 44--48; \\
M. Bordemann and J. Hoppe, \PLB 317 1993 315; \PLB 325 1994 359.

\bibitem{gschw84}
M. Green and J. Schwarz, \NPB 243 1984 285.

\bibitem{m0}
J. Hughes, J. Liu and J. Polchinski \PLB 180 1986 370.
\hskip24pt

\bibitem{m1}
E. Bergshoeff, E. Sezgin  and P. K. Townsend, \PLB 189 1987 75;\\
M. J. Duff, P. Howe, T. Inami and K. S. Stelle, \PLB 191 1987 70;\\
A. Achucarro, J. Gauntlett, K. Itoh and P.K. Townsend, \NPB 198 1987
441;\\
E. Bergshoeff, E. Sezgin, Y. Tanii and P. K. Townsend, \AoP 199 1990 340-365.

\bibitem{t1}
A. Achucarro, J. M. Evans, P. K. Townsend and D. L. Wiltshire,
\PLB 198 1987 441.

\bibitem{m}
M. Duff, {\sl Class.Quantum Grav.} {\bf 5} (1988) 189; {\sl Ibid}
{\bf 6} (1989) 1577.


\bibitem{m1a}
P. K. Townsend, \PLB 202 1988 53.

\bibitem{m1c}
A. Strominger, \NPB 343 1990 197;\\
G. T. Horowitz and A. Strominger, \NPB 360 1991 197.

\bibitem{m1d}
M. Duff and J. X. Lu, \PLB 273 1991 409; {\sl Class.Quantum Grav.}
{\bf 9} (1992) 1; \NPB 390 1993 276.

\bibitem{m2}
C. G. Callan, J. A. Harvey and A. Strominger, \NPB 359 1991 611;
\NPB 367 1991 60.

\bibitem{vzUFZh}
D. V. Volkov and A. Zheltukhin, {\sl Ukr. Fiz. Zhurnal}
{\bf 30} (1985) 809 [in Russian].

\bibitem{zhelt1}
A. Zheltukhin, {\sl Theor.Math.Phys.} {\bf 77} (1988) 377.

\bibitem{pen}
R. Penrose, {\sl J. Math. Phys.} {\bf 8} (1967) 345;
{\sl Rep. Math. Phys.} {\bf 12} (1977) 65;\\
R. Penrose and M.A.H. MacCallum, {\sl Phys.Rep.} {\bf 6} (1972) 241;\\
R. Penrose and W. Rindler, {\sl Spinors and Space--Time}, V.1,2.
Cambridge Univ. Press, 1986.

\bibitem{fer}
A. Ferber, \NPB 132 1977 55.

\bibitem{schir}
T. Schirafuji, {\sl Progr. Theor. Phys.} 70 (1983) 18.

\bibitem{twbc}
A.K.H. Bengtsson, I. Bengtsson, M. Cederwall and N. Linden, \PRD 36 1987
1766;\\
I.Bengtsson and M.Cederwall, \NPB 302 1988 104;\\
M.Cederwall, \PLB 226 1989 45.

\bibitem{twes}
Y. Eisenberg and S. Solomon, {\sl Nucl.Phys.} {\bf B309} (1988) 709.

\bibitem{twplyus}
M. S. Plyushchay, {\sl Phys.Lett.} {\bf B240} (1990) 133.

\bibitem{stv}
D. Sorokin, V. Tkach and D. V. Volkov, {\sl Mod. Phys. Lett.} {\bf A4}
(1989) 901.

\bibitem{vz}
D. V. Volkov and A. Zheltukhin, \JETPL 48 1988 61;
\LMP 17 1989 141; \NPB 335 1990 723.

\bibitem{stvz}
D. Sorokin, V. Tkach, D. V. Volkov and A. Zheltukhin, {\sl Phys. Lett.}
{\bf B216} (1989) 302.

\bibitem{sp}
D. P. Sorokin, {\sl Fortshr.Phys.} {\bf 38} (1990) 923;\\
P. S. Howe and P. K. Townsend {\sl Phys. Lett.} {\bf B259} (1991) 285;\\
F. Delduc and E. Sokatchev, {\sl Class. Quantum Grav.} {\bf 9} (1992) 361.

\bibitem{ps92}
A. Pashnev and D. Sorokin, \CQG 10 1993 625.

\bibitem{gs92}
A. Galperin and E. Sokatchev, {\sl Phys. Rev.} {\bf D46} (1992) 714.

\bibitem{hsstr}
N. Berkovits, {\sl Phys. Lett.} {\bf
232B} (1989) 184; {\bf 241B} (1990) 497; {\sl Nucl. Phys.} {\bf B350}
(1991) 193; {\bf B358} (1991) 169; {\sl Nucl. Phys.} {\bf B379} (1992) 96; {\bf
B395} (1993) 77.\\
E. Ivanov and A. Kapustnikov {\sl Phys. Lett.} {\bf B267} (1991) 175; \\

\bibitem{to}
M. Tonin, {\sl Phys. Lett.} {\bf B266} (1991) 312; \\
{\sl Int. J. Mod.Phys} {\bf 7} 1992 613; \\
S. Aoyama, P. Pasti and M. Tonin, {\sl Phys. Lett.} {\bf
B283} (1992) 213.

\bibitem{dis}
F. Delduc, E. Ivanov and E. Sokatchev, \NPB 384 1992 334.

\bibitem{dghs92}
F. Delduc, A. Galperin, P. Howe and E. Sokatchev, {\sl Phys. Rev.}
{\bf D47} (1992) 587.

\bibitem{bstv}
I. A. Bandos, D. P. Sorokin, M. Tonin and  D. V. Volkov,
{\sl Phys.Lett.} {\bf B319} (1993) 445-450.

\bibitem{bcsv}
I. A. Bandos, M. Cederwall, D. P. Sorokin and  D. V. Volkov,
 {\sl Gothenburg Preprint} {\bf ITP-94-10 (hep-th/9403181)} 10pp,
 {\sl Mod. Phys. Lett.}{\bf A} [in press].

\bibitem{gs93}
V. Chikalov and A. Pashnev, {\sl Mod.  Phys.  Lett.} {\bf
A8} (1993) 285; Preprint ICTP, 1993.

\bibitem{gs2}
A. Galperin and E. Sokatchev, \PRD 48 1993 4810.

\bibitem{tp93}
P. Pasti and M. Tonin, \NPB 418 1994 337.

\bibitem{bers94}
E. Bergshoeff and E. Sezgin, \NPB 422 1994 329.\\
E. Sezgin, Preprint CTP TAMU--58/94, November 1994; hep--th 9411055.

\bibitem{bh}
I. A. Bandos, {\sl Sov. J. Nucl. Phys.} {\bf 51} (1990) 906;
 {\sl JETP. Lett.} {\bf 52} (1990) 205.

\bibitem{bzst}
I. A. Bandos and A. A. Zheltukhin,
 {\sl Theor. Math. Phys.} {\bf 88} (1991) 358;
 {\sl JETP. Lett.} {\bf 54} (1991) 421;
 {\sl Phys. Lett.} {\bf B288} (1992) 77;
 {\sl Sov. J. Nucl. Phys.} {\bf 56} (1993) N1, [198--213 In Russian].

\bibitem{bzst1}
I. A. Bandos and A. A. Zheltukhin ,
Preprint {\sl IC/92/422} ICTP, Trieste, 1992, 81p.,
{\sl Sov. J. Elem. Part. Atom. Nucl.} [in press].

\bibitem{bzm}
I. A. Bandos and A. A. Zheltukhin ,
 {\sl JETP. Lett.} {\bf 55} (1992) 81;
 {\sl Phys. Lett.} {\bf B288} (1992) 77;
 {\sl Int. J. Mod. Phys.} {\bf A8} (1993) 1081;
 {\sl Sov. J. Nucl. Phys.} {\bf 56} (1993) n1,[198--213 In Russian].

\bibitem{bzm1}
I. A. Bandos and A. A. Zheltukhin ,
 Preprint {\bf DFPD/94/TH/35}, Padova, 1994, {\bf hep-th/9405113};
{\sl Class. Quantum Grav.} [in press].

\bibitem{bz0}
I. A. Bandos and A. A. Zheltukhin , {\sl JETP. Lett.} {\bf 51} (1990)
547; {\sl JETP. Lett.} {\bf 53} (1991) 7; {\sl Phys. Lett.} {\bf B261}
 (1991) 245; {\sl Theor. Math. Phys.} {\bf 88} (1991) 358; {\sl
 Fortschr. Phys.} {\bf 41} (1993) 619--676.

\bibitem{al}
J. De Azcarraga and J.Lukiersky, {\sl Phys. Lett.} {\bf B113} (1982)
170.

\bibitem{sig}
W.Siegel, {\sl Phys. Lett.} {\bf B128} (1983) 397.

\bibitem{gsw}
M. Green, J. Schwarz and E. Witten, Superstring Theory, CUP, 1987.

\bibitem{spinsup}
S. J. Gates Jr.  and H. Nishino, {\sl Class. Quantum
Grav.} {\bf 3} (1986) 391.\\
J.  Kowalski-Glikman, {\sl Phys.
Lett.} {\bf B180} (1986) 358.\\
R.  Brooks, F.  Muhammed and
S. J.  Gates Jr. {\sl Class.
Quantum Grav.} {\bf 3} (1986) 745.\\
R. Brooks. {\sl Phys.  Lett.} {\bf B186} (1987) 313.\\
J.  Kowalski-Glikman, J.  W. van Holten, S.  Aoyama and J.
Lukierski {\sl Phys. Lett.} {\bf B201} (1987) 487.\\
A. Kavalov and R. L.  Mkrtchyan. Spinning superparticles.  Preprint
Yer.PhI 1068(31)-88, Yerevan, 1988 (unpublished).\\
J. M. L. Fisch, \PLB 219 1989 71.

\bibitem{ghs}
A. Galperin, P. Howe and K. Stelle, \NPB 368 1992 248.

\bibitem{gds}
A. Galperin, F. Delduc and E. Sokatchev, \NPB 368 1992 143.

\bibitem{sok}
E. Sokatchev, {\sl Phys. Lett.} {\bf B169} (1987) 209;
\CQG 4 1987 237.

\bibitem{nis}
E. Nissimov, S. Pacheva, S. Solomon,
\NPB 296 1988 469; \NPB 299 1988 183; \NPB 297 1988 349;
\NPB 317 1989 344; \PLB 228 1989 181; \\
E. Nissimov and S. Pacheva, \PLB 221 1989 307.

\bibitem{kr}
R.  Kallosh and M. Rahmanov, \PLB 209 1988 233; \PLB 214 1988 549.

\bibitem{balach}
A.P. Balachandran, F. Lizzi and  G. Sparano,  {\sl Nucl.Phys.}
{\bf 263} (1986) 608.

\bibitem{dirac}
P.A.M. Dirac  {\sl Proc. Roy. Soc.} {\bf A268} (1962) 57;

\bibitem{nambu}
Y. Nambu, {\bf 1970}, {\sl Lecture on the Copenhagen Summer Symposium}.

\bibitem{polya}
A. M. Polyakov \PLB 103 1981 207.

\bibitem{howet}
P. Howe and Tucker, \JPA 10 1977 L155.

\bibitem{wieg}
P.Wiegmann, \NPB 323 1989 330.

\bibitem{gikos}
A.  Galperin, E. Ivanov, S.  Kalitzin, V. Ogievetsky and E. Sokatchev,
\CQG 1 1984 498; \CQG 2 1985 155.

\bibitem{npen}
E.T. Newman and R. Penrose, {\sl J. Math.Phys. } {\bf 3} (1962) 566


\bibitem{zup}
B. Zupnik, {\sl Sov.J. Nucl.Phys.} {\bf44} (1986) 794
[in Russian].

\bibitem{ght}
A.Galperin, P.Howe and P.K.Townsend, \NPB 402 1993 531.

\bibitem{howe94}
P. Howe \PLB {332} 1994 {61}.

\bibitem{sohnius}
M. Sohnius, \NPB 136 1978 461.

\bibitem{Dragon}
N. Dragon, {\sl Z. Phys.} {\bf C2} (1979) 29.

\bibitem{GrimWZ}
R. Grimm, J. Wess and B. Zumino, \NPB 152 (1979) 255.

\bibitem{bandzim}
I. A. Bandos and V. G. Zima, {\sl Theor. Math. Phys.}
{\bf 70} (1987) 76--91.

\bibitem{supergrav}
V. P. Akulov and  D. V. Volkov,  \JETPL 16 1972 438; \PLB 46 1973 109;\\
V. P. Akulov, D. V. Volkov and V. A. Soroka,  \JETPL 22 1975 187;\\
J. Wess and B. Zumino, \PLB 66 1977; \PLB 74 1978 51.

%\bibitem{Iv&Kr}
%E. A. Ivanov and S.O. Krivonos, {\sl Theor. Math. Phys.}  {\bf 58} (1984)
%200; {\sl Lett. Math. Phys.} {\bf 8} (1984) 39.


\bibitem{weyl}
U. Lindstr\"om and G. Theodoridis, \PLB 208 1988 407;\\
A. Karlhede and U. Lindstr\"om, \PLB 209 1988 441;\\
J. Barcelos--Neto, \PLB 245 1990 26.

\bibitem{free}
R. D'Auria and P. Fr\'e, \NPB 201 1982 101.

\bibitem{town}
P. K. Townsend, \PLB 277 1992 285.

\bibitem{bnsv}
I.Bandos, A.Nurmagambetov, D.Sorokin and D.Volkov,
{\sl Sov. JETP Lett.} (in press).

\end{thebibliography}
\end{document}